\documentclass{article}

\usepackage{amssymb}
\usepackage{graphicx}
\newcommand{\be}{\begin{equation}}
\newcommand{\ee}{\end{equation}}
\newcommand{\bea}{\begin{eqnarray}}
\newcommand{\eea}{\end{eqnarray}}

\begin{document}

\title{Mathematical General Relativity}

\author{Alan A.~Coley}

%-----------------------------------------------------------------

%%\author{Alan~Coley}
%\footnote{aac@mathstat.dal.ca}
%%\vspace{5pt}
%%{\it {: {\small {aac@mathstat.dal.ca}}}
%%, Halifax, Canada.
%\pacs{PACS}%%\pacs{98.80.Es, 98.80.Cq}

\maketitle

\begin{abstract}

We present a number of open problems within
general relativity.
After a brief introduction to some
technical mathematical issues and  the  famous singularity theorems,
we discuss  the  cosmic censorship hypothesis
and  the  Penrose inequality,
the uniqueness of black hole solutions and 
the stability of Kerr spacetime and  the  final state conjecture,
critical phenomena and  the  Einstein-Yang--Mills equations, and a
number of other problems in classical general relativity.
We then broaden  the  scope and discuss some mathematical problems
motivated by quantum gravity,  including AdS/CFT correspondence and problems in
higher dimensions and, in particular, the
instability of anti-de Sitter spacetime, and in 
cosmology, including  the  cosmological constant problem and dark energy,
the stability of de Sitter spacetime and cosmological singularities and
spikes. Finally, we briefly discuss some problems in
numerical relativity and relativistic astrophysics.

\end{abstract}

\vspace{30pt}

Alan Coley:
Department of Mathematics and Statistics, Dalhousie University,
Halifax, Nova Scotia, B3H 4R2, Canada [aac@mathstat.dal.ca].

%%============

%%one draft long--- pages with details re GR. book?where extended GR? liv. rev, arxiv, cqg
%%%Problems in GR--some in cosm/QGconsistent notation: GR throughout?

\newpage

\section{Introduction}

We discuss a number of open problems within classical
(mathematical) general relativity (GR). There have been a number of 
unsolved problems put forward
previously.       
120 open questions were presented by Yau  in \cite{Yau1982}, mainly in 
differential  (and primarily Riemannian) geometry,
but they include five problems in GR: problems Y115, 116 and 119 therein, and
problem
114 on cosmic censorship and problem
120 on  defining
the total angular momentum. These problems were also listed among the
fourteen unsolved problems in GR by
Penrose (RP12  and 10) in \cite{Yau1982}, which 
were mostly formal  questions about
null infinity, conformal properties and conservation laws, needed for  the fundamental problems and conjectures that subsequently followed.
From the assumption of cosmic censorship \cite{Penrose1969} and
some reasonable physical properties,  a 
sequence of inequalities can then be derived (e.g.,  the  Penrose inequality RP13 \cite{Penrose1973}),  the  validity of which can be interpreted as giving support for cosmic censorship.
There were a number of other problems presented, including (as yet unsolved) problems on black holes 
(e.g., RP14). In addition, many open problems in the vacuum case can be generalized to problems which include matter.

There was also a number of technical problems in mathematical GR presented by 
Bartnik \cite{Bartnik}.
Indeed, by this time GR had matured sufficiently for mathematical arguments 
to have replaced previous heuristic arguments and for rigorous problems to be posed.
Problem RB32 is  the  so-called ``Bartnik splitting
conjecture'': {\em{Let M be a ``cosmological spacetime'' satisfying  the  timelike
convergence condition: then either M is timelike geodesically incomplete or M
splits as $R \times M^3$ isometrically and thus is static}}
(see also Y115 in \cite{Yau1982}).
In  the  case of a four dimensional (4D) vacuum globally hyperbolic,
spatially compact
spacetime, if M splits it is
necessarily flat and covered by $R \times T^3$,
and thus for a non-vacuum ``cosmological
spacetime"   the  conjecture asserts that  the  
spacetime either is singular or splits. 
The  basic
Lorentzian splitting  conjecture RB32
was resolved in \cite{EJH}, and can be regarded as a singularity  theorem 
since a spacetime is time–like geodesically incomplete except in the unphysical cases in which the spacetime
splits.
The current status of  the  conjecture by Bartnik has been further reviewed in \cite{Galloway96}.
In addition, problem RB43  \cite{Bartnik} concerned  the  question of 
finding an exact  2-body  solution of  the  Einstein field equations and 
the stability of a system of two
orbiting bodies  in GR (see also \cite{Focusissue}).
This is currently important for 
the study of  the  merger of two compact objects in which  gravitational radiation
is copiously produced, especially in light 
of recent LIGO observations \cite{LIGO}. This will be briefly discussed at  the  end of this paper.

\newpage

More recently a  (selected) discussion of
advances in  the  mathematical understanding of Einstein's GR
was provided by Chrusciel, Galloway and
Pollack  (CGP) in \cite{CGP}, in which a list of 20 open problems was  presented (in Appendix A therein). 
These problems are often technical and many are specific sub-problems of those discussed in  the  main text here (where more recent progress is also described) and
include initial-boundary-value-problems such as, for example, 
to find a well posed initial boundary value problem for  the  vacuum Einstein equations that is well suited for numerical treatment (CGP7) and
problems regarding initial data (CGP16 \& 19). In addition, their problem
(CGP11) asks whether  the  existence of Cauchy horizons, not necessarily compact or analytic,
implies  the  existence of local isometries. 
Another concise survey on  the  Cauchy problem was given in \cite{AretakisRodnianski},
a number of problems in mathematical GR were discussed in 
\cite{Coleyquestions}, and
a survey on strong cosmic censorship is anticipated soon  by Dafermos and Luk.

\newpage
\section{Open problems in General Relativity}

There are many problems concerning the  general  mathematical properties of solutions of Einstein's field equations of GR. 
A sufficiently differentiable
4D Lorentz manifold is considered. The  causal structure of the manifold
is determined by the
Lorentzian metric, which 
satisfies  the  Einstein
field equations. These equations form a 
hyperbolic system of  quasi-linear partial differential
equations which are generally coupled to additional partial differential equations governing
the matter in the spacetime \cite{Rendall2002}.
First the  vacuum case is investigated (with a Ricci
flat metric). It is important to study  the  Cauchy problem; here
the unknown variables in  the vacuum constraint equations, which consist of a spatial metric
and a symmetric 3-tensor belonging to a  Riemannian 3D submanifold (and any appropriate initial data for 
the matter fields), are taken as the  initial data for  the vacuum evolution
field equations.

The Einstein field equations are covariant (that is, 
invariant under an arbitrary change of coordinates), which further
complicates  the their formulution \cite{LARS99}.
The resulting vacuum field equations are not, in  the normal sense, hyperbolic. However,  
in so-called harmonic coordinates
the vacuum
equations {\em{do}} form a quasi-linear hyperbolic
system and consequently  the  Cauchy problem is well posed and
the usual local existence results follow \cite{CB69}. In addition,  the  constraints and
gauge conditions are  are preserved by  the  evolution equations.
In particular,  the  global regularity 
for small, smooth initial data (with suitable asymptotic decay) in 3D
for a coupled Klein-Gordon-Wave system   
(which is a simple analogue of  the  Klein-Gordon system in GR) was investigated in \cite{Ionescu}.
Results for  the field equations of GR coupled to a variety of different 
material fields are also known; these matter fields include not only perfect fluids and 
gases satisfying kinetic theory, but also scalar
fields, Maxwell fields and Yang-Mills fields, 
and even exotic matter motivated by string theory \cite{Narita},
 and various combinations thereof. 
Perfect fluid results
generally only  apply when the energy density of the fluid is  bounded (uniformly)
\cite{Rendall2002}.

%%\newpage

\paragraph{Local existence:}
The standard theorem states that given infinitely differentiable  ($C^{\infty}$) 
initial data for  the  vacuum Einstein field equations there is, defined at least for a finite time period,
a smooth solution of  the  equations which
corresponds to this initial data \cite{CB80}.
The local existence of solutions of  the field equations in GR is 
reasonably well understood. 
The basic global uniqueness theorem then guarantees the uniqueness
of the  maximal development \cite{CB69} of any such initial Cauchy
data (up to diffeomorphisms that fix  the  initial Cauchy hypersurface, and that 
the  subsequent long term solution depends continuously on  the  initial data \cite{CB80}.
Unfortunately, proving general global existence theorems
is very difficult and currently not possible using standard mathematical techniques \cite{Rendall2002}. 
The  conformal
method \cite{CB80} is often used for investigating the  Einstein field equations.  
In this case the free data are specified and
the  resulting constraints reduce to 4 elliptic equations.
When the mean curvature 
is constant (see later), the constraint equations simplify even further to a linear system of 3 (elliptic) equations and a decoupled non-linear, scalar  equation (the Lichnerowicz equation).

Due to the conformally invariant
nature of the causal structure in a Lorentzian manifold,
the  compactified ``regular conformal field equations''
were constructed from  the  Einstein field
equations, 
which constitute a first order symmetric hyperbolic system and consequently leads to
well posed evolution equations and, in addition, 
global existence
theorems for small data which follow from  stability results for quasi-linear hyperbolic equations.
In particular,  Friedrich \cite{Friedrich86} proved a  future global existence result for 
small initial data that is close to  conventional data specified on a Minkowski hyperboloid.
The conditions on generic initial data specified on an
asymptotically flat Cauchy surface that subsequently leads to a Cauchy development with regular
conformal completion is still not known. 
Friedrich has, however, pioneered an
approach in which  the  conformal structure at spatial
infinity is investigated (\cite{Dain,Kroon}, and
new problems with regularity have been pointed out; (also see  
\cite{HansRingstrom17,Ringstrom2015} and references within).

Currently the  only known global existence theorems
for  the  full 4 dimensional  (with no symmetries) field equations in GR   are  
(i) the non-linear stability theorem of Minkowski space \cite{ChristodoulouKlainerman90},  
(ii) the  
semi–-global existence result  for  the  initial value problem on a  hyperboloid \cite{Friedrich86} and 
(iii) the  semi-–global existence
result for spatially compact spacetimes with a hyperbolic Cauchy surface 
\cite{AnderssonMoncrief}. All of these results are for small data.
It is known that vacuum (and electrovac) spacetimes
containing a Cauchy horizon which is analytic and
``ruled by closed null geodesics'' admit a non-trivial
Killing vector field \cite{Isenberg85}. There are many results
possible when special spacetimes with symmetries are considered, and these results
are briefly summarized below. Of course, a spacetime admitting a Killing vector
fields is not generic, but such results may be interpreted in favour of a 
strong cosmic censorship theorem (see below).

%%\newpage

\paragraph{Cases with symmetry:}
A  global existence theorem is possible in special spacetimes with
symmetry  \cite{Rendall2002,LARS99}.
A well known example is the 
global existence theorem for (smooth) spherically symmetric static solutions
with perfect fluid or  collisionless matter  \cite{Rendall2002}.
The  symmetry of a spacetime is characterized by  the total number
of Killing vector fields and their nature.
Let us consider spacetimes with an $r$–-dimensional Lie
algebra of space–like Killing vector fields. Then, for each $r$  (with $r \leq 3$), there are a number of
results on global existence and various conjectures regarding cosmic censorship \cite{LARS99}.                     
In particular, when $r$ = 3 (Bianchi models \cite{Chrusciel55}) and the special polarized Gowdy models with
$r=2$ (see references below),  the  global behaviour of solutions of  the  Einstein field
equations is reasonably well understood.

There are only partial
global existence results in the  general case of
$r = 2$ (with a local $U(1) \times U(1)$ $G_2$ symmetry). In addition,
there is no resolution of the  cosmic censorship conjecture \cite{LARS99}.
The original global existence theorem for Gowdy spacetimes
with $R \times T^3$  topology was presented in  \cite{Moncrief81}.
Subsequently this theorem was generalized to spacetimes on $S^3$
and $S^2 \times S^1$  \cite{Chrusciel47} (a subclass of metrics that are not ``generic'' remain to be investigated).  Theorems on  vacuum Gowdy spacetimes with a global constant mean curvature foliation
were first proven in \cite{IsenbergMoncrief1982}.
Cosmic censorship has also been investigated 
utilizing  the  asymptotic behaviour of curvature invariants in the polarized Gowdy spacetimes \cite{Chrusciel54} and more generality  \cite{Kichenassamy7}. Unfortunately,
the  horizon  structure and extensions in  polarized Gowdy spacetimes is
extremely complex \cite{Chrusciel49,Chrusciel53}.  There are no large data
global existence results nor theorems on cosmic censorship 
in  the $r$ = 1 case ($U(1)$ symmetry) or $r$ = 0 (no symmetry) case. However, 
from numerical experimentation
there are some conjectures on  the
generic behaviour in  $U(1)$ spacetimes as well as small data semi–-global existence results in  the  direction   of expansion \cite{Choquet01,Choquet03}.

%%\newpage 

\paragraph{Differentiability:}

Technical
questions relating to differentiability are
important for many of the mathematical problems described, and especially 
for well-posedness \cite{Rendall2002}. The
smoothness  of  the  Cauchy data for
the initial value problem of a system of partial differential equations 
and that of  the  corresponding solutions
are determined by  the partial differential  equations themselves. 
Indeed, there is a correspondence between  the differentiability of
the free data and  the  full data in the
Einstein constraint equations.

But there is important motivation for assuming weaker differentiability conditions  than the
natural condition of $C^{\infty}$. For example, one reason is that
the matter fields are not necessarily $C^{\infty}$ (so that results may not be applicable
in physically important scenarios). Another
reason is the potential for 
continuing local existence results of lower differentiability solutions  to global existence results. 
In addition, there are also
indications that generally Cauchy horizons may not be differentiable
\cite{Chrusciel52}.

Therefore, it is important to find a theoretical framework for  
investigating  the  constraint and evolution 
equations for spacetimes with milder differentiability (e.g., to prove results with
lower differentiability; e.g., 
to consider
metrics of regularity class $C^{1,1}$  \cite{Kunzinger}
(where  the  first derivatives of  the  metric functions are locally Lipschitz continuous functions,
which is perhaps a more appropriate class than the regularity class $C^2$
in many  situations of physically interest).
The initial data for  the  vacuum field equations can be relaxed 
further and assumed to belong to a local Sobolev space, especially in existence and uniqueness theorems.  
Indeed, in harmonic
coordinates  the vacuum field equations constitute a quasi-linear hyperbolic system of equations,
and the usual theorems imply that  the resulting Cauchy problem is well-posed in
an appropriate Sobolev space \cite{HughesKato} (see also \cite{KR,Tataru}).

\paragraph{Singularity theorems:}

The singularity theorems are an immense theoretical
accomplishment within GR and, indeed, within mathematical
physics in general \cite{SenovillaGarfinkle}.
Penrose's original singularity theorem \cite{Penrose65} introduced the  concepts of geodesic incompleteness
(wherein geodesic curves exist
that cannot be extended in a regular manner to the complete  spacetime)
to characterize singularities, Cauchy
hypersurfaces and global hyperbolicity, and closed trapped surfaces \cite{Penrose1979}. These concepts 
have proven crucial to the development of mathematical GR. Subsequently,
Hawking understood that closed trapped
surfaces may occur in a Universe that was expanding to the past, which could then also 
lead to an initial cosmological singularity  \cite{Hawking1966}. 
This then led to  the famous singularity Hawking and
Penrose theorem  \cite{PenroseHawking}:
{\em{if a convergence and a generic condition
holds for causal vectors, and there are no closed time-like curves and there exists at least one
of  the  following:
a closed achronal imbedded hypersurface,
a closed trapped surface,
a point with re-converging light cone,
then  the  spacetime has incomplete causal geodesics.}} From this
theorem the  discovery of  the  cosmic microwave background then implies that 
a singularity may of actually occurred in  the  past in our own Universe \cite{HawkingEllis}.

The singularity theorems of  Penrose and Hawking show that, under reasonably general conditions,
spacetime singularities may be inevitable
\cite{Penrose65,PenroseHawking}. However, these theorems do not say very much about  
the  properties of generic
singularities. There are also ``general'' spacetimes that do not have singularities
\cite{Senovilla2012}; e.g., 
the analysis in the  Penrose singularity theorem does not
necessarily imply that a trapped surface will form during evolution. However,
it was subsequently  proven  \cite{Christodoulou2009} that in
vacuum spacetimes a trapped surface can dynamically form from regular
initial data that does not have a trapped surface (this theorem was generalized in
\cite{Klainerman2014,Klainerman2012}). A 
sequence of marginally outer trapped surfaces, in which the area approaches zero and hence forms 
an apparent horizon,  were then constructed
within a region up to  the  ``center" of gravitational collapse in  the  full 4D vacuum
GR field equations \cite{AnLuk}. 
The existence of marginally outer trapped surfaces are important for proving  the  positive mass theorem and
the Penrose inequality \cite{AnderssonMetzger} (see below).

There are many outstanding
questions. These include obtaining more general
singularity theorems with both  weaker  energy conditions and with weaker assumptions of differentiability, and 
investigating whether there is a  relationship
between geodesic incompleteness and the divergence of
curvature invariants \cite{Senovilla2012}.
In addition, there are various related outstanding problems in mathematical 
cosmology. Generic spacelike singularities are often called cosmological singularities. 
But oscillatory singularities may also correspond to
the spacelike part of generic black hole singularities \cite{SenovillaGarfinkle,Senovilla2012}, since
the  mass inflationary instability at  the  inner horizon of a
rotating black hole that is accreting may be followed by an oscillatory collapse to a spacelike singularity 
\cite{Hamilton}. There are also  the  questions of whether the
singularities of GR can be  resolved by quantum effects and  whether singularity theorems are  possible in higher dimensional spacetimes
\cite{Coleyquestions}. But perhaps  the  most pressing unresolved question within GR concerns the cosmic censorship hypothesis.

\newpage

\subsection{Cosmic censorship}

The theorems of Penrose and Hawking \cite{Penrose65,PenroseHawking} imply  the  existence of singularities.  
However, although  the simple Schwarzschild
spacetime does  contain a singularity, the singularity is hidden inside a black hole event
horizon and is thus not visible from the exterior region. Therefore, it is of interest to ask
whether the gravitational collapse of physical matter leads to singularities like the
Schwarzschild  singularity in the sense that they are hidden inside an event horizon and they are not time-like (satisfying  weak cosmic censorship and strong cosmic censorship, respectively \cite{Penrose1979}).

As a result, Penrose introduced  \cite{Penrose1969}  the  cosmic censorship hypothesis. This
asserts, roughly speaking, for the gravitational field equations of GR
coupled to ``physical"  matter fields, that 
no ``naked singularity" will develop from regular
``general and realistic" initial Cauchy data. Essentially
a naked singularity is a singularity
from which light can escape to distant regions. Such a singularity poses
severe problems, since  the  future is no longer predictable \cite{HawkingEllis}.
It is not reasonable to conjecture that such singularities can never occur, since
there are examples of spacetimes that admit a naked singularity. However, such spacetimes are 
highly symmetric and it is not inconceivable that naked singularities formed
under small (general, physically realistic) perturbations will be hidden by an event horizon.  
In fact, analysis implies that naked singularities only form in spacetimes with  symmetry \cite{IsenbergMoncrief1982}.

For example, naked singularities are known to exist in Taub-NUT spacetime \cite{MisnerTaub,Chrusciel53}.
In addition, so-called ``shell
crossing"  will occur in a pressure-less fluid or ``dust" 
spherically symmetric Lemaitre-Tolman-Bondi-de Sitter collapse \cite{Goncalves},
in which a central locally naked singularity will form
from typical regular initial data and where various
curvature scalars diverge.
The most rigorous theorems are by Christodoulou \cite{Christodoulou94,Christodoulou99b},
who proved that naked singularities can develop from
regular initial, asymptotically flat data for solutions of  the  globally inhomogeneous spherically symmetric 
Einstein field equations coupled to a massless scalar field.
However, perhaps  the  work 
on critical behaviour by
Choptuik \cite{Choptuik2015} (see later) is more important.

Therefore, it is desirable to formulate cosmic censorship as a precise mathematical
conjecture, and then seek a proof.  The  
formulation of  the  conjectures by Christodoulou \cite{Christod99} are perhaps 
the most widely utilized,
although   the  revised formulation of strong cosmic censorship given in the
introduction of \cite{Christodoulou2009} corrects some flaws \cite{LukREF}
in  the  original formulation given in \cite{Christod99}.
Unfortunately, such theorems, which lie within  the  
global theory of partial differential equations, are very
difficult to prove  \cite{Ringstrom2010}. A simpler approach might be,
since there are no time-like singularities in a globally hyperbolic spacetime, to attempt to
to formulate (strong) cosmic censorship by asserting that spacetime must be globally hyperbolic under 
appropriate general
conditions. Unfortunately, a maximal
Cauchy development of initial data in a globally hyperbolic spacetime
may not constitute the  complete spacetime. An alternative approach is to seek 
a counterexample to any mathematical cosmic censorship
conjecture.

There are two other particular issues.
First, it is very difficult to  mathematically define a naked singularity 
since  the field equations are essentially hyperbolic and hence extending a solution to regions which are in causal contact with  the  singularity is problematic; consequently, we seek an alternative definition
of a naked singularity that can be formulated
mathematically. 
The second problem is genericity. 
It is  not possible to prove a general theorem that 
naked singularities cannot exist since there
are exceptional spacetimes which,
for any reasonable definition, contain a 
singularity in which the  maximal development is, in fact, extendible.
Therefore a ``genericity condition'' is needed, otherwise such a statement of
cosmic censorship would necessarily fail.
However, we are ultimately interested in  the physical process of
gravitational collapse, and so we want to be careful not to formulate a 
conjecture that might admit an artificial counterexample.  Consequently, we wish to
formulate  the cosmic censorship conjecture so as to avoid
counterexamples are not physical, but at the same time not at the expense of making cosmic censorship impossible to disprove \cite{Rendall2002}.

There are, in fact, both weak and strong
cosmic censorship hypotheses, and these two  hypotheses are only marginally 
related.  The  weak cosmic censorship hypothesis states that:
{\em{For generic initial data to  the  evolution problem in GR, there cannot be naked singularities.}}
It is of prime importance to prove  the  weak cosmic censorship conjecture.
This is such an open problem that  the  correct formulation of  the  statement is not even known  \cite{Christod99}. This conjecture is discussed extensively in \cite{Wald1998}.
The work of Christodoulou discussed above \cite{Christod} essentially  proves weak cosmic censorship for  asymptotically flat, spherically
symmetric Einstein–-scalar field spacetimes. In addition, Christodoulou
also presented examples of naked singularities that develop from initial  Cauchy
data  \cite{Christodoulou99b}).

The strong cosmic censorship hypothesis
states that:
{\em {A generic solution to Einstein's field equations cannot be continued beyond  the  Cauchy horizon.}}
For reviews of this version of the cosmic censorship conjecture see \cite{Isenberg92,Chrusciel49}.
It would be informative to prove the two versions of cosmic censorship even for
solutions of the vacuum  field equations  or,
more generally, for classes of spacetimes with symmetry. In
particular, it is of interest to 
prove that  the  maximal
Cauchy development for generic vacuum data sets  is equal to  the  maximal vacuum
extension of a 3D compact manifold.

%%\newpage

Alternatively, and as noted above, a different approach  
is to look 
for a counterexample  to cosmic censorship. If a variety of classes of possible counterexamples can be shown to not to work, it might even be viewed as
evidence for  the  likely veracity of  the  conjecture. 
Recently, a plausible vacuum counterexample to cosmic censorship with a negative cosmological constant has been proposed utilizing  the  superradiant instability of  the  Kerr-–AdS black hole \cite{Dias2015}.
Another plausible 4D asymptotically anti--de Sitter counterexample 
modelling holographically an electrically charged localised defect
within
Einstein-–Maxwell was presented in
\cite{Horowitz2016}. In addition, this counterexample is generic \cite{Bizon}, 
unlike  the  spherically symmetric collapse candidates
which are finely tuned.

\newpage

\paragraph{The Penrose inequality}

The analysis used in  the singularity
theorems has been utlizied in several other problems in GR. For example, it has been applied in  the  positive
mass theorem  \cite{SchoenYau} and 
the  study of  rigidity in asymptotically flat spacetimes with non-negative 
scalar curvature. It was demonstrated by Penrose \cite{Penrose1973} that a spacetime obtained by
evolving  initial data will contain a naked singularity if a  particular inequality is not satisfied.
This inequality, called the {\em{Penrose inequality}},
relates  the  area of the so-called apparent horizon
and  the ADM mass within this marginally (outer) future-trapped surface.
Consequently, initial data
that does not satisfy this inequality could be used to construct a counterexample to the weak cosmic
censorship conjecture. On the other hand,
a ``proof'' of this inequality migth constitute evidence in favor of
this form of cosmic censorship and might, in addition,
lead to an analytical approach to attack  the  cosmic censorship conjecture
\cite{Moncrief81E,LARS99}). An important open problem is to
prove  the  Penrose inequality or construct a  generic counterexample.

Recently
the Riemannian analogue of  the  Penrose inequality was proven \cite{Huisken}. Unfortunately, there is no such
proof in  the  Lorentzian case, even in  simpler spacetimes with symmetry.
For example, in spacetimes with spherical symmetry only a version of the conjecture using the
energy rather than  the ADM  mass has been shown to be true. A number of proofs have been presented
under a variety of different further assumptions such as, for example,  the  existence of 
constant mean curvature foliations  \cite{Moncrief96}) and global conditions on  the  spacetime  \cite{Bray}.
A formulation of  the  conjecture for certain null surfaces in spacetimes (the {\em{null Penrose conjecture}}) was studied in \cite{NPC}.

In fact, the  Penrose inequality is but one of a large number of mass inequalities for spacetimes
\cite{DainS}. E.g., a similar inequality exists utilizing the  Penrose
quasi-local mass \cite{Tod}. In addition, it is of interest to  generalize
these  mass inequalities to   non-time-symmetric initial data.
There are also stronger versions of  the  Penrose inequality that involve angular
momentum and/or electric charge (or even a non-zero cosmological constant). All of these generalizations 
lead to additional open problems \cite{SenovillaGarfinkle}.
There are also additional refinements of some of the  conjectures. For example, the Gibbons-Penrose inequality, 
gives a stronger lower bound if multiple black holes are present \cite{Gibbons}. There is also
Thorne's hoop conjecture 
\cite{Thorne1972}, which has been very useful \cite{Wald1998}. This inequality attempts to exploit
the reasonable notion that   
the energy/matter content of a physical black hole, which is an extremely localized object,
must be severely restricted in all spatial directions. A
rigorous
reformulation of  the  hoop conjecture was recently given \cite{Senovilla2008}.

\newpage

\paragraph{Critical phenomena:}
Choptuik  \cite{MWChoptuik} first studied  the  spherically
symmetric  gravitational collapse
of a massless scalar field in GR numerically and found strong evidence for a universal, self-similar (critical)
solution at  the  threshold of black hole formation. 
In particular,  a one 
parameter  family of solutions to  the  coupled
Einstein-matter equations was considered.
The continuous
parameter $p$ was used to
prescribe the initial data which, in turn, 
controls  the  strength of  the matter which is initially undergoing gravitational collapse. 
For small values  of $p$, where the strength of gravity is not appreciable,  the  
spacetime remains regular everywhere (e.g., in the case of a massless scalar field
the radiation will disperse to infinity).
For large values of $p$, gravity is sufficiently strong  that
some of  the  matter becomes trapped within a singular black hole.
For a special ``critical'' value of $p$, there exists a  solution which is self similar and 
corresponds to  the  black hole formation  threshold. 
Current studies
suggest that almost every model containing black hole collapse
will admit critical behaviour \cite{Choptuik2015}.
It is an important mathematical problem within
GR to better 
understand these critical solutions and  
to describe the  corresponding critical behaviour in gravitational collapse
in  the general case in which there are no symmetries.
Much work has now been done on critical phenomena in various models  \cite{Choptuik2015} including,
for example, in Yang--Mills theory (see below).
In \cite{Reiter}  the  authors claimed to have proven  the  existence of
Choptuik's solution using computer assisted methods.

\subsection{The Einstein-Yang-Mills equations}

Yang-Mills (YM) fields  over, for example, $ R^4$ or $S^4$ 
are studied using techniques in partial differential equations.
A number of important
questions, including their  mathematical
stability and existence proofs, have not yet  been resolved.
A fundemental  property of YM theory  is  its invariance
under an infinite-dimensional
group.  The YM field equations depend on the ``gauge'' choice of a
section of  the  Lie algebra valued bundle; in an appropriate gauge the
YM equations become a quasi-linear elliptic system.
SInce bundles, connections and curvature
play a fundamental role in YM theory, it is also of
interest to mathematicians \cite{Yau1982}.
Numerical computations indicate that  static, spherically
symmetric Einstein-Yang-Mills (EYM) equations have non-singular, asymptotically flat
solutions \cite{BartnikMckinnon}.

In 4D there are smooth solutions of the
YM equations globally for reasonably general initial conditions. 
Assuming sufficiently differentiable initial data,
global existence
was proven in  Minkowski spacetime \cite{Eardley}.
Subsequently, local existence for data of finite energy  was proven \cite{Klainerman95},
and due to energy conservation this  implies a global existence result.
Global existence  for 1+3 dimensional, globally hyperbolic
manifolds was proven \cite{Chrusciel56,Kichenassamy7}).
A  global existence proof (to the  future) for
small initial data on a hyperboloid in Minkowski
spacetime, which was
subsequently generalized to Maxwell and YM matter, 
was presented in \cite{Friedrich91}. 
In addition, asymptotically flat, spherically symmetric 
``particle-like'' solutions of  the  coupled EYM equations (with gauge group SU(2)  and
a regular interior)  are known.
However,  their properties are not understood fully \cite{Olinyk}.

YM theory has also been investigated in higher 
dimensions $D$. For $D \geq 5$  solutions exist in which
singularities develop in a finite time, although
computations provide evidence this is
stable in that the blow-up  occurs for an open set of initial data.
In addition, 
there are indications that this blow-up is separated
from dispersion by a critical self-similar solution. 
However, there is no rigorous proof of blow-up, even in 5D.
And in 6D, apparently different
singularities (from those in 5D) can form \cite{LARS99}.

{\em{Wave maps}} with values on spheres have also been studied, since
they capture effects found in YM theory but in two less dimensions
where  theorems are easier
to establish. In particular, a solution
with  the  expected properties of  the  critical solution for wave maps 
in 4D in which a singularity forms was proven to exist \cite{Bizonwavemaps} (see below).
A  global
existence theorem for  the  classical  wave map equation is an  important unresolved problem. 
There are small data global existence results for the  non-linear ($\sigma$-–model, hyperbolic
harmonic) wave map equation in space dimensions $n \geq 2$. 
Unfortunately,  global existence results for “large data” are  only known for solutions with symmetry.
Even global existence for  the  wave map
equation  in $n = 2$ has not been proven. However,  utilizing energy estimates,
global existence can be proven in  the  case $n = 1$,
\cite{LARS99,AnderssonGudapatiSzeftel}. It is perhaps of interest to study the
U(1) symmetric vacuum case in 4D  in  the  presence of a space–like Killing vector field
which is hypersurface orthogonal.
The Einstein field equations then simplify to
3D gravitational equations which are coupled to wave map matter, and so this case
interpolates in difficulty between  the  4D Einstein field equations
and  the Gowdy equations which are highly symmetric \cite{SterbenzTataru}.

In particular, for YM theory in $n\geq 5$ and wave maps in $n\geq 3$ (from $D=n+1$
dimensional Minkowski spacetime into the $n$-sphere) there are explicit
self-similar solutions which are examples of finite-time
singularity formation from regular initial data \cite{BizonBiernat}
(which are linearly
stable; non-linear stability was proven later in \cite{Donninger}).  The
{\em{critical solutions}} sitting at  the  borderline between global
regularity and blowup are self-similar and have exactly one
unstable direction.  Wave maps (from 3+1 Minkowski spacetime to
the 3-sphere) are perhaps a particularly  good toy model to study critical behaviour in
gravitational collapse \cite{Choptuik2015}, although the
critical solutions in this case are continuously (rather than discretely)
self-similar.
The dimensions $n=4$ for YM theory and $n=2$ for wave maps are critical and in these
(and only these) dimensions there exist stationary solutions
(the harmonic maps).  The
energies of these stationary solutions determine  the  thresholds for
singularity formation (that is, solutions starting from initial data
with lower energy are globally regular).  It is also known that the
process of singularity formation in these particular dimensions proceeds
via shrinking of  the  stationary solution to zero (along the
symmetry orbit generated by scaling).  Note that in  the  case of
wave maps  the  geometry of  the  target is important.  For example,
for negative curvature targets  the  wave map equation in $n=2$ has no
stationary solution and consequently  the  evolution is globally
regular. Later, new critical solutions were discovered in the
EYM model (which
evade Birkhoff's theorem and which already have rich dynamics even in the
spherically symmetric case) \cite{Chmaj}.

%%\newpage

\paragraph{Collisionless matter models:}
 
Alternative forms of matter such as,
e.g.,  self-gravitating
collisionless matter have also been considered  \cite{LARS99,Rendall2002,Andreasson}.
Global existence and uniqueness  theorems for
sufficiently smooth solutions of 
the  classical Boltzmann (and Vlasov-Poisson) equations  
in Newtonian theory have been extended to GR.
For example, the
global existence of weak solutions and 
convergence of classical solutions with close to equilibrium
initial data have been proven. And there are
existence theorems for spacetimes with symmetry, such as static spherically symmetric, 
plane and  hyperbolically symmetric spacetimes, and
a subset of the general Gowdy spacetimes. Spacetimes undergoing 
spherically symmetric collapse have also been investigated.
It is known that models with Collisionless matter
can evolve  globally without singularities. In fact,
many models can be isotropic
to the future.
The general case has been investigated 
numerically \cite{Rendall2002,LARS99}, but 
no analytical results are known.
An open problem is to prove a global existence result for general  (spatially inhomogeneous) collisionless matter models for small initial data and, in addition, to prove an existence and
uniqueness theorem  for spatially homogeneous Einstein-Vlasov models in the
large initial data case.

%%%%%%%%%%%%%%

\newpage

\subsection{Uniqueness and stability}

The proof of stability  involves the mathematical study of the  asymptotic behaviour of solutions of the  highly non-linear Einstein system of partial differential equations, which is 
notoriously very hard. However, there are  some special spacetimes of particular interest for
which  proofs exist. We note that 
de Sitter and anti-de-Sitter spacetimes will be discussed later.

\paragraph{Stability of Minkowski spacetime:}

Minkowski spacetime is known to be globally stable \cite{Christodoulou93}.  
The global existence (for small data) and  the  stability of Minkowski spacetime 
under  the Einstein field equations was first investigated in  \cite{ChristodoulouKlainerman90,Christodoulou93}, where it was proven that if asymptotically flat initial data close to those of Minkowski spacetime
are specified on a hyperplane for  the  vacuum equations, then  the  maximal Cauchy development of this initial data is geodesically complete. The  asymptotic properties
of  such solutions have also been discussed. In particular,  
the possible behaviour  in
a suitable neighbourhood of infinity of asymptotically flat spacetimes,
in which  the  initial matter distribution has compact support, has been investigated
There have been a number of recent generalizations of these results
\cite{Bieri}.
The stability of Minkowski space for Einstein-Vlasov systems was studied in \cite{EinsteinVlasov}.

\paragraph{Uniqueness of black holes:}
If  the  final asymptotic state of any spacetime is conjectured to be either Minkowski spacetime or a black hole spacetime, then it is of interest to determine whether a given black hole 
is  the  unique stationary solution. The  question of the uniqueness of non-vacuum  steady state  black hole spacetimes  goes under  the  colloquial name of ``no-hair" theorems. 
Neither of these problems have been completely resolved.  
However, the uniqueness of black hole spacetimes with additional symmetry,  such as axial symmetry or rotational symmetry, has been established.
The  uniqueness of  the  GR  4D Schwarzschild and Kerr solutions was first was discussed in \cite{HawkingEllis},
and a theorem for Schwarzschild spacetime was established in \cite{Bunting,Israel}.
The Kerr spacetime is the unique stationary regular vacuum solution 
(subject to some particular conditions)
\cite{Kerr}, and its
uniqueness was proven in \cite{Carter,Robinson}.
The  uniqueness of  the non-vacuum  rotating electrically charged
Kerr-Newman  black hole solution has not yet been proven in general  \cite{Newman65,Mazur}.
In the case that a black hole spacetime is
real analytic, then it is known to be unique.
It is anticipated that this result is also true when
regularity is relaxed to the assumption of infinite differentiability,
but there are only some partial results known in this case. 
In particular, for a given stationary black hole spacetime there are no other approximate  (under small perturbations)  stationary solutions other than the given 
black hole solution itself. In addition, 
no other stationary exteriors are permitted
when certain special structures on  the  event horizon
are postulated.

\subsection{Stability of Kerr spacetime}

If the Kerr-Newman black hole spacetimes do indeed represent  the  unique stationary states in GR, 
then a very important problem is to prove their stability (particularly under small perturbations). 
The stability of  the  Kerr metric was reviewed 
in  \cite{Chand} and discussed  in \cite{Heusler}.
A particular goal is to demonstrate that
perturbations of the
Kerr  solution (and, indeed, the Schwarzschild  solution \cite{Holzegel1016}) exponentially decay.
Recently there has been substantial and exciting progress made  in  the  linearised problem \cite{DHR}.
However,  results for  the  full non-linear problem are still elusive.

A full, quantitative (i.e., which provides precise decay estimates) proof of  the  linear
stability of Schwarzschild spacetime has recently been established by Dafermos, Holzegel
and Rodnianksi in \cite{DHR}.
In  the  passage from  the  linear to  the   non-linear stability of Schwarzschild 
new difficulties need to be overcome, some of which are similar to those encountered in  the  stability of Minkowski \cite{Christodoulou93}.
The non-linear stability of  the  Schwarzschild spacetime under  the  restricted class of non-trivial
axially symmetric polarized perturbations was proven in \cite{KlainermanSzeftel}. 
The new ideas introduced therein may be applicable in a more general setting and will hopefully  be an important first step to proving  the  non-linear stability of Kerr to gravitational
perturbations.
Indeed, proving  the  non-linear stability of Kerr 
has become one of  the  primary areas of
mathematical work in GR \cite{Shlapentokh}
(for further background see \cite{lecturenotes} and  the  important references
to Dafermos and others cited within; perhaps  the  most concise and up to date survey
can be found in  the  introduction of \cite{recentpaper}).

Current observations are consistent with  the  predictions of GR,
suggesting that  the  end point of mergers is a Kerr
black hole. Indeed, all numerical computations provide support in favour of the non-linear stability  
of  Kerr (and Kerr-Newman) black holes (at least
for certain ranges for  their  angular momentum)
\cite{Zilho}.  
However, there is still no complete mathematical analysis of  the  stability of
the generic Kerr black hole, nor a comprehensive  understanding  of
its dynamics under arbitrary non-linear perturbations. A rigorous
proof of  the  Friedman instability \cite{Friedman} (that for a real analytic, stationary and asymptotically
flat spacetime with a non-empty ergoregion and no future event
horizon,  the  energy of certain solutions diverge  as time increases) was provided in \cite{Moschidis16} 
in  the  more
general case of smooth but not necessarily globally real analytic spacetimes
(only a unique
local continuation condition was imposed, which
always holds if  the  manifold is locally analytic but
can also be inferred in  the  case when  the  manifold possesses an appropriate second
Killing vector field). In contrast,
Aretakis \cite{Aretakis} has proven that axisymmetric extremal horizons 
(such as, for example,  the  extremal Kerr-Newman  spacetimes)
are unstable under
linear scalar perturbations, unlike the
subextremal case for which decay is known for all derivatives along  the  event horizon.

\newpage

The  stability analysis has been extended to include a non-zero cosmological constant.
Recently there have been results regarding non-linear perturbations in Kerr-de Sitter
black hole spacetimes
(at least in  the  slowly rotating case) \cite{Hintz}.
However, an investigation in the case of a negative cosmological constant is much more 
difficult due to the potential superradiance instability of
the Kerr-AdS black hole itself \cite{Holzegel}). We shall discuss 
the stability of both the  de--Sitter spacetime and the anti--de--Sitter spacetime below.
An investigation of  the  stability of models which include matter is also important, particularly in  the  cosmological context (see later). Unfortunately, generalizations to even include a simple inhomogeneous perfect fluid are very difficult mathematically
due to  the  possible formation of shocks and (in  the  case of dust) shell-crossings.
It would first be  necessary to  further
develop appropriate
criteria to detect, using classical hydrodynamics, the possible development of shocks.

\newpage

\paragraph{The Final State Conjecture:}

The {\em{final state conjecture}}, originally discussed by
Penrose in~\cite{Yau1982}, is  the  statement that generic
asymptotically flat and vacuum initial data 
(which can include black holes, and hence technically also matter if it's
confined to the interior of the black holes, but no naked singularities)
evolve to a solution which either
disperses~(in which case there are no black holes) or else
eventually asymptotes (in the exterior) to finitely many Kerr
solutions (a collection of unbound black~holes)
moving away from each other.
That is, for small perturbations of Kerr, not only is null infinity still complete, 
but in the exterior region the Kerr~metric
is asymptotically stable~(up to a redefinition of the parameters).
The final state conjecture includes the statements of the non-linear stability
of the (exterior of) Kerr spacetime~\cite{DafermosHolzegelRodnianski} and  
weak cosmic censorship as special cases.
It is of importance to
prove this conjecture, but it
is unlikely that this will be done in the near future
\cite{LukREF}.

%%%%%%%%%%%%%%%

If the  metric is real analytic, as in the original formulation of  the  {\em{no-hair}} theorem, then
the assumption of closeness~to~Kerr in  the  stationary uniqueness theorem is no longer necessary~\cite{HawkingEllis}. 
However, this assumption of analyticity is not well motivated, since
stationarity does not guarantee that the Killing vector field is timelike everywhere in the exterior~(e.g., in
the Kerr geometry itself, in the  so-called~ergoregion outside  the  black and white holes  the Killing vector field becomes spacelike), and analyticity only follows from  the  ellipticity of  the  reduced equations in this case.  
If the analyticity assumption were weakened to smoothness only, then the
consequent uniqueness property would support the  more ambitious Penrose
final state conjecture for vacuum spacetimes arising
from generic asymptotically 
flat Cauchy data (not necessarily initially close to Kerr), whereby
the  Kerr exterior would not only be a stable endstate, but it would also be the
general endstate for all non-dispersing vacuum solutions.

%%%%%%%%%%%%%%%

The final state conjecture includes not only  the  non-linear stability of Kerr 
but also  the  evolution of (generic) arbitrarily large
amplitude gravitational waves.  
Consequently, by seeking
quasi-normal modes in black hole mergers (i.e., with
{\em{no hair}}) in gravitational wave observations 
the dynamics of GR
and hence  the  final state conjecture is investigated.  
Therefore, this very
important open question in
mathematical GR is directly related to
gravitational wave observations.
Some evidence in favor of the
final state conjecture has been provided by numerical 
simulations \cite{YangPaschalidis}.

\newpage

\subsection{Other problems}

\paragraph{Newtonian limit:}
A precise mathematical statement that within GR the
limit of a divergent speed of light 
leads to Newton's theory of gravity is problematic.
A formulation of a suitable 
definition of the  Newtonianian limit of GR with physically
reasonable properties was first presented by  Ehlers \cite{Ehlers}. 
Unfortunately, it is not clear  that there are general families
of solutions of  the field equations within GR which have a Newtonian limit with
the definition adopted.  In addition, 
a mathematical justification of  the ``post-Newtonian'' approximation relies on 
the existence of suitably
differentiable families of such solutions.
In particular, Bartnik's problem
RB21 \cite{Bartnik} seeks a proof that solutions of  the  Einstein field equations reduce to 
solutions of Newton's theory in a  rigorously defined limit . 
And problem RB20 \cite{Bartnik} addresses the question of the 
range of validity of both the post-Newtonian and the post-Minkowskian asymptotic expansions.
There has been much work done 
on finding solutions in which the metric
satisfies  the  Einstein field equations approximately, 
consisting of both 
numerical computations and analytical techniques such as asymptotic
expansions, linearisation techniques and matching analyses.

\paragraph{Geodesic hypothesis:} An important postulate of GR is that
the intrinsic geometrically defined spacetime
geodesics are the trajectories along which  
point particles with negligible mass move.
It remains to rigorously
prove that such test particles  move on 
geodesics in spacetime.
This problem (RB23) was considered by Einstein himself and,
despite extensive investigations
using asymptotic expansions \cite{JEhlers},
has not yet been fully resolved.
An essential obstacle is a rigorous definition of the 
``negligible mass limit".
In addition,  within GR a physical object itself causes 
a gravitational backreaction  in  the  spacetime due its own motion.

\paragraph{Constant mean curvature:}
There are a number of problems in mathematical GR,  involving both  the  initial data constraints
and the evolution equations, in
which the results obtained
rely upon the  underlying spacetime admitting a constant mean curvature (CMC) Cauchy surface 
(or slice). However, not all
spacetimes have CMC slices \cite{BartnikPollack}
(although the known examples are very
special with a high level of symmetry), and this has
important ramifications.
Consequently, perhaps the  most serious question concerning CMC slices is their generality.
If spacetimes generally admit CMC
slices then  they can be used for proving generic properties, and the
special examples where they do not exist can be neglected.
In  \cite{DiltsHolst},
the known results were reviewed and a suite of conjectures concerning  the  existence of CMC slices and their generality were made.

\newpage

\paragraph{Scalar curvature invariants:}
It is known that a 4D 
(and higher dimensional \cite{CHP2}) Lorentzian spacetime can
be completely characterized by  its  scalar curvature invariants (built polynomially from
the Riemann tensor and its covariant derivatives), unless it
belongs to  the   special  class of ``degenerate Kundt'' spacetimes \cite{CHP}. 
The special class  of Kundt spacetimes
are those that admit a geodesic, expansion-free,
shear-free, twist-free null vector.
Therefore,  in general,
a Lorentzian spacetime is completely locally characterized by its scalar  polynomial curvature invariants.
A  Riemannian manifold is always characterized by such scalar
invariants.

The  ``inverse'' problem of determining
when a spacetime manifold can be explicitly constructed from its scalar curvature invariants is of interest.
It is also important to find
the minimal set of such invariants.
In particular, to
determine  the  minimal set of scalar curvature invariants
needed for the explicit characterization of
a non-degenerate 4D Kundt spacetime.
We note that  the  Petrov type of  the  Weyl tensor  in 4D  is partially
characterized by its scalar polynomial curvature invariants \cite{krameretal}.

\paragraph{Horizon evolution:}  
There has been a lot of work done in trying to formulate an appropriate definition of  the  ``boundary of a black hole". If  the  evolutions along the two future
null vector field directions normal to 
a closed oriented space-like 2-surface (isomorphic to $S^2$) in a spacetime 
are both area-non-increasing to  the future
(or one of  the  null mean curvatures
is zero), then that 2-surface  is said to be future
trapped (or, respectively, an ``apparent horizon'').
It is a particularly important open problem to formulate an appropriate
definition (and hence a local characterization)
of a dynamical horizon for an evolving black hole.
A lot of  the analysis of
the  evolution of apparent horizons is in the linear regime.
This implies that to first order the  horizons do not move.  
The  non-linear  evolution of apparent horizons is still not understood.
Recently scalar curvature invariants have been utilized to 
identify and locate (``geometric'') horizons \cite{ColeyMcNutt}.

\paragraph{Gravitational wave memory:}
The study of asymptotia, including that in 
non-asymptotically flat spacetimes  (such as
de Sitter spacetime) is also of importance. In particular, such issues arise in  the  problem 
of gravitational wave memory.
GR describes gravitational radiation from astrophysical phenomena \cite{BGYone}.
Gravitational wave memory is a permanent displacement
of a gravitational wave detector after  the  wave
has passed, and has been extended to  the  full non-linear theory
of GR by Christodoulou \cite{Christodoulou1991}.
It is known  that  the  memory found 
in an (ordinary)  linearized situation (due to fields that do not reach null infinity) and (null memory)
in  the  non-linear theory  (due to fields that do reach null infinity) are two different
effects.  The  ordinary
memory is very small, whereas  the  null memory
is sufficiently large to be detected by Advanced LIGO and
LISA as a permanent displacement of test masses, and also in radio telescopes like NANOGrav as a change in  the   pulse frequency of pulsars. 
It is also known that electromagnetic fields and neutrino radiation enlarge  the  memory effect \cite{BGYone}. 

Much of  the  work treats gravitational memory
in an asymptotically flat spacetime. However, we appear to live in
an expanding Universe, not an asymptotically flat spacetime.
Moreover,  the  sources of gravitational waves are
rare and those that have been detected to date 
have been at distances at which  the  expansion of the
Universe cannot be neglected. As  the  detectors become
even more sensitive,  detections due to
sources at even greater distances might be anticipated in which  the  expansion of
the Universe will be even more important.
A proper treatment of memory in an expanding Universe
is thus crucial. 
Gravitational wave memory in  the  case where  the  sources and detector are in an expanding ${\Lambda}$CDM cosmology was examined in \cite{BGYtwo}. 
The  Universe is highly inhomogeneous.
However, for simplicity, a de Sitter cosmological spacetime was assumed and  the  gravitatational radiation was treated in  the  short wavelength approximation.
Results very similar to those of gravitational wave memory in an asymptotically flat spacetime were found,   except that  the  total magnitude of  the  memory effect is amplified by a redshift-dependent factor and can be affected by gravitatational lensing.

\paragraph{Alternative theories of gravity:}
Many of the problems discussed here can be rephrased in  the  context of alternative (to GR) gravity theories.
Such problems are outside the scope of the present article, but
would include several questions within  cosmology (see later and \cite{Ishak}).
For example, it might be of interest to study the  stability of  the  isotropic Friedmann model \cite{Goode} on  approach to an initial cosmological singularity in alternative  theories of  gravity.
In particular, this question has been investigated in gravity theories with higher--order 
curvature terms in the  Einstein-Hilbert action \cite{Middleton}.

\newpage

\section{Broadening  the  scope}

\subsection{Quantum gravity}

The attempt to formulate a  fundamental theory which unites 
the standard model of particle physics, which includes 
the three forces of 
electromagnetism and  the  strong and weak nuclear forces,
together with
gravity (i.e., GR)  
into a single, fully consistent, unified theory
of quantum gravity (QG), is a primary goal of
theoretical physics.

GR is modified by QG effects, and new
gravitational physics are possible. However, 
it seems that such modifications do not dramaticallly influence the  macroscopic
behaviour of stellar systems and black holes over astrophysical timescales \cite{Barcelo}.
Perhaps  the  most spectacular consequence of  the  unification of
GR and quantum mechanics is that 
a black hole 
evaporates due to the  emission of Hawking radiation \cite{Hawking}, but 
even here the classical properties of the black hole are not affected significantly.
However, attempts to unify quantum phenomena and GR, such as string theory or loop QG,
have motivated a number of new theoretical problems.

\paragraph{AdS/CFT correspondence:}
The  unique maximally symmetric Lorentzian spacetime 
with constant negative scalar curvature in any dimension
is Anti-de Sitter (AdS) spacetime. Such spacetimes are important in theories of QG formulated in terms of string theory or modern extensions
such as M-theory. 
In fact, AdS spacetimes currently
are of primary importance in theoretical physics due to  the
conjectured  Maldacena AdS/CFT correspondence. This  
is a hypothetical fundamental  equivalence between string theory on
an asymptotically AdS spacetime and a  conformal field
theory (CFT) on its  boundary \cite{mald,Klebanov}.
CFT are conformally invariant  quantum field theories describing elementary particles,
and include  theories of Yang–-Mills type.

Gauge/gravity duality postulates that a  quantum mechanical force such as electromagnetism 
(or forces arising from the  weak or  strong interactions) can be described by a string theory in an AdS spacetime with an additional dimension.
Since  AdS/CFT correspondence implies a non-perturbative formulation of string theory (with particular boundary conditions), it could 
potentially lead to significant advances in our comprehension of string theory and QG.
And, of course,  the  utility of this gauge/gravity duality is that 
strongly coupled CFT can consequently be investigated by considering the  corresponding
weakly interacting theory of  gravity, which may be more tractable  mathematically. 
A number of problems in nuclear and condensed matter physics 
(including  the  modeling of
heavy ion non-equilibrium collisions) have
been investigated 
by translating those problems  into mathematical problems in string theory.
Namely, a AdS/CFT dictionary is utilized to translate 
a strongly coupled
CFT to  its  string dual. The corresponding classical AdS gravitational theory is then investigated, and 
potentially useful results in  the  CFT are then deduced. 
The down side of this holographic approach is that
the gravitational theory  in  the  non-stationary sector is not always understood.

\subsection{Instability of Anti-de Sitter spacetime}

The AdS/CFT correspondence
provides strong motivation for studying asymptotically AdS
spacetimes. Of course,  the  dynamics of AdS
spacetimes  within classical GR is itself of interest.
Minkowski and de-Sitter spacetimes \cite{HolzegelShao} were proven to be non-linearly stable many years ago
\cite{Christodoulou93,Friedrich}.
However, this is not the case for AdS spacetime and, in fact, 
it has very recently  been postulated that  the  AdS spacetime is  not stable
to (arbitrarily) small perturbations  \cite{PiotrBizon},  
By instability we mean that 
for large classes of initial data any perturbation of AdS spacetime, no matter how small, leads to  black hole formation.

Prior to  the  work of \cite{Bizon2015},
there was no real conjecture of instability of AdS
(however, see \cite{Dafermos}). 
In particular, the
note \cite{PiotrBizon} simply observes a
physically obvious fact (proved in \cite{Anderson}) that
a perturbed AdS spacetime cannot settle down to a globally regular
stationary asymptotically AdS spacetime because  the  only such
spacetime is AdS itself (and  the  energy is conserved for reflecting
boundary conditions).  However, this is no argument for instability
(and even less for black-hole formation) because  the  evolution can
be a globally-regular-in-time quasi-periodic perturbation of AdS  \cite{Anderson}.  Perhaps the
most important subsequent result was  the  perturbative construction
of time-periodic asymptotically AdS solutions 
(with negative cosmological constant in an $n+1$ dimensional spherically symmetric spacetime) and numerical
evidence of their instability \cite{MalRost}.

The study of  the  global
non-linear stability of AdS was given great impetus by the recent work of Bizon
and Rostworowski \cite{Bizon} 
in  an  analysis of a spherically
symmetric  massless scalar field within GR.
It was suggested by numerics that AdS is non-linearly unstable to a weakly
turbulent mechanism. Arbitrarily small black holes can form. The 
mass of such black holes seems to be determined
by  the  initial data. 
And although this
non-linear instability appears to be present for a variety of typical perturbations, 
there are many perturbations 
for which such an instability
is not generated  \cite{DiasSantos}. This consequently 
appears to suggest the
existence of islands of stability \cite{DiasGreen}. 
Indeed, understanding which configurations lead to instability and 
those that do not is a crucial question, and many 
analytical and numerical tools have been deployed to disentangle stable from unstable initial data
in  the  search for islands of stability  \cite{Martinon}.

\newpage

In particular, secular growth and non-linearities were shown to
occur that can then lead to resonances and weak turbulent behaviour \cite{Bizon}. 
Heuristically, this  turbulence
is believed to be triggered by the  generation of secular terms by resonant four-wave interactions
which drive  the  instability.
Perturbations to third order in  the  amplitude of  the  linear seed were used in the analysis.
Modifications of standard perturbation theory, 
such as  the  resonant approximation \cite{Bizon2015}, 
can then be used to represent the  dynamics for
particular time scales. However, these approximations depend on   
spherical symmetry. It
is not currently known if any GR solution with a fully resonant
spectrum must lead to a non-linear instability; however, it appears that such a  condition is necessary
for  the  existence of  such a  weakly turbulent instability. 
The role of  the  fully
resonant spectrum remains unclear \cite{DiasSantos}; but some results do not rely on any
resonances \cite{Moschidis}. It is of interest to determine
if  the  non-dispersive character of  the  linearized spectrum
is necessary for the
turbulent instability. It is also of interest to investigate the generality of  such an
instability.

It is of great interest to investigate this further without the assumption of spherical
symmetry. A number of different seeds have been used in 
calculations to
third order in  perturbation theory  \cite{DiasSantos,DiasGreen}, and it appears
that  the  gravitational possibilities are richer than in the  spherically symmetry case studied
previously \cite{Bizon}. 
It is, of course,  essential to investigate  the  non-linear stability of AdS without any
assumptions on symmetry.
However, despite
some recent research \cite{Martinon}, such an investigation 
is currently not possible, neither from an analytical nor even a numerical standpoint.
It is therefore of primary importance to determine  the possible end point
of instability for non-spherically symmetric perturbations 
for  AdS spacetimes in any dimension \cite{PiotrBizon}. That is, to
determine 
if  the  conjectured non-linear AdS instability 
has different properties  in more general spacetimes than those in
spherically symmetric scalar field collapse. And, in addition, to determine
whether this  conjectured non-linear instability leads to a weakly 
turbulent mechanism that subsequently develops a ``cascade'' towards higher frequencies and the formation
of black holes. 
This is  discussed in more detail in  \cite{Jalmuzna} (also see  the  review  \cite{Martinon}).
We remark that non-linear instability was recently
proven for  the Einstein-massless Vlasov system in the case of spherical symmetry
\cite{Moschidis}.

\newpage
\subsection{Higher dimensions:}

Additional spatial dimensions (to the
usual three of normal space) are utilized in string theory \cite{string}. 
Higher dimensional spacetimes are also considered in a number of  scenarios within cosmology. 
If, indeed, higher spatial dimensions exist, then fundamental questions on the  character
of  the  internal space of the Universe arise.

The investigation of black holes in higher dimensional GR is
particularly important.
Classically gravity, and especially  black holes, has a much richer structure in higher dimensions than in standard 4D. In particular, it is known that  the stationary Kerr 
black hole is  unique in 4D GR.  However, there are are a variety of
asymptotically flat,  exact vacuum black hole
solutions in higher dimensions \cite{EmparanReall}.  
An important open question is to determine  the  uniqueness and stability of black holes in higher dimensions.
This leads to a  number of interesting new problems,
including the reformulation of the  hypotheses of analyticity, non-degeneracy and connectivity in the
study of uniqueness of black-holes in higher dimensions.
It is also of interest to 
classify all vacuum ``near-horizon'' geometries with compact cross-sections.

Differential geometry has recently been further developed
in higher dimensions \cite{CHP,higherdimensions}. 
In particular, 
despite  the  fact that  originally the  singularity theorems were investigated in 4D, 
there exist results for co-dimension
two closed trapped surfaces in any arbitrary dimension. In fact, the 
definition of being trapped can be extended to submanifolds of any co-dimension  
(although a suitable condition on the curvature must be postulated in order to guarantee that focal points to  the  submanifold exist).
In particular, co-dimension three
closed trapped submanifolds
in higher dimensions were studied in \cite{Galloway}. And recently, 
a positive mass theorem has  been presented in arbitary dimensions \cite{SchoenYau2017}.

It is also of interest to investigate the question of stability in higher dimensions.
However, this is a 
much more complicated question since while
the radiative decay of solutions is greater in higher dimensions, which would 
enhance stability, 
there is also a larger number of degrees of freedom, which, in general, would increase  the  potential for an instability. We note that
numerical evidence does suggest that 
a number of higher dimensional black holes are unstable \cite{EmparanReall}.

Cosmic censorship in higher dimensions is a very difficult problem and, indeed, is not even a well posed
problem. In fact,
there is a lot of numerical evidence that
higher dimensional black holes can be unstable under
gravitational perturbations, so that cosmic censorship consequently fails \cite{LehnerPretorius2010}.
This was originally demonstrated 
for black strings and p-branes \cite{GregoryLaflamme} 
(in 4D such an
instability is not possible).
There is numerical 
evidence  \cite{LehnerPretorius2010} that black strings are unstable and ``pinch off'' in a finite
time, thereby resulting in a singularity that is naked.
This  constitutes a violation of  weak cosmic censorship
because the  initial data is not finely tuned. 
However, these 
spacetimes have compact additional spatial dimensions. The
black rings of \cite{EmparanReall2002} also suffer from 
this and various other types of instabilities
\cite{BRI}.

\newpage

Rapidly spinning 
black holes 
deform significantly  and are not, in fact, dissimilar to black branes. 
Myers-Perry (MP) black holes \cite{MP}  (which are
higher-dimensional analogues of Kerr  black holes)
with large angular momenta have been shown to exhibit an
``ultraspinning'' instability when the 
spacetime dimension exceeds six \cite{EmparanMyers}. 
In particular,
the final point of  the instability of  asymptotically flat axisymmetric  6D MP
black holes was investigated and the  ultraspinning instability  was shown to
produce a ``sequence
of concentric rings which are connected by segments of black membrane on  the  rotation plane'' \cite{Figueras}.
In  the  non-linear region these concentric rings
become ever more thin, and in a finite time  a naked singularity forms
resulting in  a violation of 
weak cosmic censorship.

%%%%%%%%%%%%%%%%%%%%%%%%%%%%

There is also interest in theoretical results such as, for example,  the  dynamical stability of higher dimensional cosmological models. Spatially homogeneous cosmologies in higher dimensions and,  
in particular, extensions of  the  BKL analysis have been investigated
\cite{Henneaux}.

\newpage

\subsection{Cosmology:} 

Cosmology is  concerned with  the  large scale behaviour of  the  Universe within  GR (or, in some cases
of interest,
a particular alternative  theory of gravity).
There are a number of problems in mathematical or theoretical cosmology of current importance.

Cosmological inflation is generally accepted to resolve  the  {\em{horizon problem}},
that  the Universe seems to be more uniform  than expected on large scales. However,
it should be noted that
alternative  explanations for this might be possible. The are many theoretical problems
with inflation, which include the
precise nature of
the hypothetical inflaton field and the  details of 
{\em{cosmic inflation}}. 
In particular, does the  amplification of quantum-mechanical fluctuations
lead to self-sustaining inflation and is it consequently still occurring in  
various places in  the  Universe.
In addition, 
it is anticipated that 
initial conditions in both the inflaton field and the geometry contain inhomogeneities.
But inflation does not  necessarily occur for
random initial conditions. For example, 
it has been demonstrated that small
field inflation is sensitive to inhomogeneities  \cite{infl}. However, 
large field inflation appears to be more robust to inhomogeneous and
anisotropic initial conditions, particularly in the  simple case that initially
any large energy gradients in  the inflaton field are confined to  the  inflationary portion of its
potential. An open problem is to determine 
the  generality of inflation for generic initial data.

Additional questions include, for example,  the  beginning and late time behaviour of the
Universe, and especially  whether  the  Universe is evolving towards a final singularity.
Using a modified theory of gravity (i.e., an alternative to GR) or invoking 
non-standard material fields,
it has been shown that  the final  or initial singularity can be replaced classically
by a {\em{bounce}} to another separate expanding region \cite{Brandenberger,EKp}.
Therefore, it can be asked whether  spacetime  evolves towards a ``big'' bounce and, indeed,  
whether it is  part of an
infinite cyclic Universe.

Cosmology is concerned with  the large scale dynamical behaviour of  the  Universe 
in which small-scale structures (e.g., those of galaxies) are negligible.
Therefore, it is usually assumed that
the  ``Cosmological Principle'' is satisfied. This principle asserts that 
{\em{on large scales  the  Universe can be accurately modeled by an  isotropic
and spatially homogeneous 
solution to the Einstein field equations.}} 
Namely, this states that a preferential
cosmological time can be chosen so that at all times, and on large scales,
space appears the same in all directions (isotropy) and at all places (spatial homogeneity).
The Universe is obviously not spatially homogeneous on the scales of galaxies or clusters of galaxies.
Consequently it would be desirable if  the  cosmological principle 
(that is, late time spatial homogenization and isotropization)
could be deduced from the field equations of GR 
under suitable physical conditions and for appropriate initial  data, rather than postulated a priori. 
This is perhaps addressed within  inflation, at least to some extent.

\paragraph{The cosmological constant problem and dark energy:}
The  most important problems in physical cosmology concern dark matter and dark energy \cite{AL42}.
These forms of matter and energy 
cannot be directly observed and
can  only be identified by their gravitational effects.  
The existence of dark matter in  the  Universe 
is deduced from  observed galaxy rotation curves,
nucleosynthesis analyses and structure formation computations \cite{Freese}. 
The form of  the  missing  dark matter
is not yet known. In fact, it is not even known
if it is a particle or
whether  the   dark
matter phenomena can be  described by a generalization of GR. 
It is, however, believed that this problem will be resolved by standard physics. 
The dark energy problem, on  the  other hand,  
is regarded as a primary impediment
to further understanding within theoretical physics
\cite{Witten2001,Steinhardt}.

The {\em{cosmological constant  problem}} has been  discussed in detail by
Weinberg \cite{Weinberg1989}. Conventional quantum field theory
(QFT) leads to an exceedingly large vacuum energy density. 
However,  the  GR equivalence principle
asserts that all forms of energy density have the same gravitational effect; hence it
is assumed that  the  vacuum energy gravitates in exactly the same way 
as a cosmological constant,
and consequently has a huge
effect on  spacetime  curvature.
But the  observed effective
cosmological constant is exceedingly small compared with
the  predicted value from QFT, so that a putative ``bare'' cosmological
constant would be necessary to
cancel out the enormous  $10^{120}$ contribution from the
vacuum.
This is an unbelievably difficult fine-tuning problem. And it gets
worse still when higher loop corrections are taken into consideration
\cite{Padilla}.
Weinberg and other physicists have advocated
that, out of all of the possible suggestions to resolve this problem, perhaps the  most reasonable is  the  controversial argument of an 
anthropic bound  \cite{Weinberg1987}.

Moreover,  for  the  last few billion years the  expansion of the  Universe has been increasing  \cite{Riess}. 
The reason for this acceleration
is commonly referred to as {\em{dark energy}}
within the standard cosmological model,
which has similar dynamical properties to a relatively small cosmological constant. 
The so-called coincidence problem of explaining why the
cosmological constant has
such a tiny observed value of
the order of magnitude of  the current 
matter  density in  the  Universe must also be addressed.
It has been conjectured that
dark energy is not a simple cosmological constant but is, rather,  a dynamical field due to,
for example, quintessence or phantom energy. 
Some researchers have also put forward 
modifications to gravity on cosmological scales
to explain the dynamical gravitational effects attributed to dark energy.
Another  possible explanation proposed for the  observed acceleration of  the  Universe 
is backreaction from inhomogeneities, that occur when averaging Einstein's field equations
on large scales.

\subsection{Averaging:}

The (effective) dynamical gravitational equations on cosmological  scales 
are presumably derived from an averaging
or coarse graining  of the  Einstein field equations. 
Averaging in cosmology is a problem of considerable  interest, and its resolution is necessary for the correct
interpretation of cosmological observations. 
Although  the  so-called standard  $\Lambda$CDM cosmological model, which is a
spatially homogeneous and isotropic Friedmann--Lemaitre--Robertson--Walker (FLRW) spacetime with a cosmological constant $\Lambda$, 
makes predictions in
exceptional agreement with  current cosmological  observations
(except for a relatively small  number of possible tensions \cite{tension}), 
it does require a dark energy that dominates the dynamics of  the  present Universe 
which, as noted above,
has never been detected directly.
Moreover,  the current Universe is neither isotropic
nor spatially homogeneous  locally. Indeed,
observations indicate that  the current  structure of  the Universe 
is very complicated. Groups and clusters of galaxies of varying size
form  the  largest gravitationally bound structures. These clusters themselves then
``form knots, filaments and sheets that thread and surround very underdense
voids''. An enormous ``cosmic web'' is consequently created \cite{web}.
An important open question in cosmology is whether averaging of inhomogeneities can lead to significant backreaction effects on very large scales.

A crucial first step is, of course, to formulate a rigorous mathematical definition of averaging in GR.
Any such spacetime volume averaging procedure  must necessarily be both well-posed and covariant,
and this introduces some fundamental new issues within differential geometry. 
The mathematical problem of formally averaging tensors and other geometric objects
on a differential manifold has recently received 
renewed attention \cite{Av,Averaging}. 
The field equations obtained by averaging
(or, more precisely, coarse graining)
do not necessarily even have the  same mathematical
structure as  the underlying Einstein field equations. 
For example, in the
macroscopic gravity approach to spacetime averaging \cite{Averaging},
which utilizes the Lie-dragging of averaging regions and
is both fully covariant and gauge invariant, the structure of the 
averaged spacetime is not necessarily even Riemannian.
However, the
macroscopic gravity approach does lead, in general  \cite{Av}, to
well-defined local averaged geometrical objects on an arbitrary 
Riemannian manifold. In particular, the  averaged structure
equations of the underlying geometry yields an appropriate set of structure equations for  the  averaged
(macroscopic) geometry. 
In addition, macroscopic gravity
gives the form  for  the additional terms (the so-called correlation functions)
arising from the averaging of  the  non-linear field equations
that necessarily occur in  the  macroscopic field equations  \cite{Averaging}.

\newpage
\subsection{Stability of de Sitter spacetime}

Rigorous mathematical results are  possible in  the  study of  the  {\em{stability of cosmological solutions}}.
This involves  an investigation of  the  behaviour of a complex set of partial differential equations at late times about a particular cosmological solution (and there are a number of exact  solutions that are of special cosmological interest \cite{AnderssonMoncrief,LAndersson14}). 
First, we recall  that if
the cosmological constant is zero and  assuming that the  matter fields satisfy  the  standard energy
conditions, 
then Bianchi type IX spacetimes recollapse and are consequently not
expanding to the future for all times.
Formally this result is encapsulated within the
“closed universe recollapse conjecture” \cite{BarrowTipler}, which
was proven in  \cite{LinWald}.
However, in the case of a non-zero positive 
cosmological constant, Bianchi type IX spacetimes do not necessarily recollapse to the future.  
In particular,  the question of the stability of de Sitter spacetime is of primary importance.

A stability theorem for de Sitter spacetime that utilized regular conformal field equations
was proven in \cite{Friedrich1986}. 
Indeed,  Friedrich proved
that all (vacuum) initial data on a regular Cauchy hypersurface sufficiently close
(in an appropriate Sobolev topology) to the initial data
corresponding to a de Sitter spacetime with a positive cosmological constant
have maximal geodesically complete Cauchy developments.
In addition, the  analysis by Friedrich also presented details
of  the resulting asymptotic behaviour of the solutions.
Consequently, de Sitter spacetime acts as  an attractor for expanding cosmological
models with a positive cosmological constant, and   
the  proof may be viewed as justification
of  the  `` cosmic no hair'' conjecture (at least in  the  vacuum case;
see also \cite{Ringstrom2015} and references within).

A cosmic no hair theorem for spatially homogeneous cosmological models
was proven by Wald \cite{Wald83}.
This general theorem, that shows that an expanding anisotropic cosmology 
which does not subsequently   recollapse 
asymptotes to an isotropic de Sitter spacetime to the future,  
does not depend on  the particular nature of
the matter fields present other than that they satisfy
the  strong and dominant energy conditions.
It can also be demonstrated that
initially expanding
solutions for initial data induced by  the  Einstein field equations with a positive cosmological constant
and coupled to various reasonable matter models exist globally in time  \cite{Rendall95}.
In particular, global results are possible for inflationary scalar
field cosmological models with an exponential self potential in which  the  rate of 
expansion increases polynomially with cosmological time \cite{Coleybook,exppot}.
We note that inflationary
behaviour also occurs for scalar field models with a power law self
potential, but this arises at intermediate times and not  to  the asymptotic future.
Local dynamics are mainly studied numerically, 
and it is difficult to establish rigorous local results.
 It is an important open problem to prove a cosmic no--hair theorem 
in generic  spatially inhomogeneous spacetimes.
Some partial results are known  for spatially inhomogeneous models with a positive cosmological constant \cite{Jensen} and, to a lesser extent,
for scalar field cosmologies with an exponential self potential  \cite{Coleybook}.

\newpage

\subsection{Cosmological singularities}

Although standard theorems inform us that singularities 
generically occur within GR, 
they give very few details about the character of these singularities  \cite{Senovilla2012}.
Belinskii, Khalatnikov and Lifshitz (BKL) \cite{art:LK63} 
have postulated that  the  approach to
a spacelike singularity to the past is ``vacuum~~dominated,~~local,~~and~~oscillatory'' (or ``mixmaster'')
for general inhomogeneous  GR cosmological models.
In particular, BKL conjectured that
due to  the  non-linearity of the
Einstein equations, if  the  matter fields do not include a massless scalar field
or its dynamical equivalent
then in a neighbourhood of the initial cosmological singularity all of the matter fields can be neglected relative to  the  dynamical anisotropy in the field equations.
The  assumptions  and associated  dynamics of  BKL have been demonstrated
to be compatible with  the field equations of GR. However, 
that does not necessarily imply that the dynamics are  generally valid in situations
of  physical interest.
Numerical simulations do support the validity of the  BKL conjecture, at least for 
particular classes of spacetimes \cite{Berger,DavidG}.

There have been a number of different approaches to study  the general structure of  cosmological singularities.
These include the more heuristic BKL metric and  Hamiltonian approaches and  the mathematically more
rigorous dynamical systems approach \cite{WE}.
In  the  dynamical
systems approach  the
Einstein field equations (not necessarily with any symmetries) are reformulated as a scale invariant asymptotically regularized first order system of autonomous ordinary (or partial) differential
equations valid in  the  approach towards a generic spacelike cosmological  singularity
\cite{Uggla03}. 
In more detail, in this reformulation  the Einstein field equations split
into a decoupled equation for  a conformal factor and a coupled system of
equations for  the associated dimensionless conformal metric,
where the conformal factor is related to the expansion of the
normal congruence to an  assumed foliation of spacelike surfaces
in a sufficiently small neighborhood
of a generic ``simultaneous'' spacelike singularity \cite{Heinzle,Uggla03}.
This approach has facilitated
a detailed description of  the  general attractor and led to precisely formulated mathematical conjectures
regarding the general asymptotic dynamical behaviour towards an initial singularity.
The reformulation also provides
a suitable foundation
for  the  numerical investigation of generic cosmological singularities \cite{LARS99}.

Only a few rigorous mathematical results had been proven until quite recently. 
Following earlier analysis \cite{Rendall},  Ringstrom presented important results on  the  asymptotic  dynamics of spatially homogeneous cosmological models  within GR \cite{Ringstrom}. Indeed,
Ringstrom proved theorems concerning the
oscillatory behaviour of generic initial spacelike cosmological singularities for models of Bianchi types VIII and IX. In particular, 
Ringstrom demonstrated that  the  past attractor
in a Bianchi type IX spacetime is contained within a   union of
vacuum subsets of   Bianchi
type I and Bianchi type II.
Unfortunately, 
the attractor is not fully specified by this theorem \cite{Heinzle}.
In addition, the results do not completely determine
the relevance of the  Kasner map  for the dynamics asymptotic
to  the  initial singularity in the Bianchi type VIII and IX  models  \cite{Brehm}.

The  emphasis to date has been on proving mathematical theorems  
for spatially homogeneous cosmological models.  
The dynamics of Bianchi type IX models are still not fully understood. In  \cite{Heinzle}
a summary of the asymptotic dynamical behaviour of Bianchi  type IX models
that has actually been proven  was presented
(also see \cite{Brehm}).
For example, arguments concerning the possible chaotic behaviour at a generic spacelike
singularity rely on the fact that the  Kasner map actually describes the  asymptotic dynamics of the
Einstein field equations. 
However, although it is
plausible that  the  Kasner map,
which is indeed associated with chaos, plays such a role,
this has not yet been proven \cite{Heinzle}.
There remain a number of outstanding questions, which include rigorously establishing the
role of Bianchi type IX models in generic cosmological singularities
\cite{Heinzle09,Uggla03,Heinzle}.
In particular, an important open problem
is to prove that  the  past attractor of  the  Bianchi type IX dynamical
system coincides exactly with  (rather than being a subset of) the  Mixmaster attractor  \cite{Heinzle,Ringstrom}. Based upon a
numerical and  qualitative analysis \cite{Hewitt2003},
the exceptional Bianchi type VI$_{-\frac{1}{9}}$ class B model
(which is of the same generality as  the  most general  Bianchi type VIII and Bianchi type 
IX class A models) also has an oscillatory
initial singularity.

In cosmological models with matter,
the BKL oscillatory dynamics have primarily been investigated in perfect fluid models with a simple linear equation of state relating the fluid pressure and density.
However, some other matter fields can have an important dynamical effect in a neighbourhood of the
initial singularity. 
A (massless) scalar field (or, equivalently, a stiff perfect fluid) will lead to  the complex oscillatory dynamical behaviour being
replaced by simple monotonic dynamical behaviour on approach to  the spacelike singularity \cite{BGZK}.  
On the other hand, 
an electromagnetic field can
produce oscillations  which are not present in vacuum or
perfect fluid  models of  the same type (e.g., models of Bianchi
types I and VI$_0$)
\cite{LeBlanc}.  And oscillatory behaviour
can also occur in all  tilting perfect fluid spatially homogeneous cosmologies
\cite{Hewitt2001,Hervik07}.

It is very important to investigate
generic initial spacelike singularities in spatially inhomogeneous cosmological models and,
in particular, to
prove  the  BKL  locality conjecture. 
Using Hubble (or expansion)-normalized variables,
both qualitative and numerical evidence provides support for  BKL oscillatory dynamics, at least for an open set of time lines \cite{Uggla13}. 
More generally, a  physical argument to justify asymptotic locality is that
extremely strong gravitational effects will lead to
particle horizons shrinking to vanishing size as  the initial singularity is approached along an
individual time line, so that
communication between different time lines may not be possible (``asymptotic silence'')
in  the  asymptotic limit.
In order to gain additional insight into general cosmological singularities in spatially inhomogeneous
spacetimes, $G_2$-models with two
commuting spacelike Killing vector fields  have been studied. 
Oscillatory BKL dynamics has been discussed in
general vacuum, spatially compact $U(1) \times U(1)$
spacetimes with zero twist  and in generic polarized $U(1)$ spacetimes 
\cite{LARS99}, and in twisting $U(1) \times U(1)$ vacuum Gowdy models on $T^3$, 
and on $S^2 \times S^1$, $S^3$
and lens spaces \cite{Maier}.
The attempt to describe  the dynamics in the  asymptotic approach towards a
generic cosmological singularity in terms of an attractor has led to a number of precise
mathematical conjectures being formulated \cite{Heinzle09,Uggla03}; indeed,  the  existence of a possible
finite dimensional attractor in  the  infinite dimensional state space has been suggested \cite{Temam}.

A  Hamiltonian approach 
(related to the BKL approach) has  been utilized to heuristically investigate  the  dynamical behaviour
of  the  Einstein-dilaton-p-form system in  the asymptotic  approach to a spacelike
singularity \cite{Damour}.  
The  limiting dynamics of  the system was described  in terms of a ``billiard''
motion in a subset of hyperbolic space bounded by  ``walls'' (this dynamical behaviour was dubbed ``cosmological billiards'').
A remarkable mathematical connection
between  the  resulting asymptotic dynamics and Kac-Moody algebras was found \cite{Damour}.
The relationship between  this approach to studying spatially inhomogeneous cosmologies 
and  the  more rigorous dynamical systems approach was presented in  \cite{Heinzle09}.
The  fermionic sector  (including classical gravitinos) of supergravity theories was also recently studied \cite{Kleinschmidt}.  A  quantum
extension of   fermionic cosmological billiards was investigated.
\cite{DamourSpindel}. 
The resulting   dynamics was described 
in term of a quantized supersymmetric
Bianchi type IX cosmology in 4D supergravity. 
Once again Kac-Moody structures were revealed.

A stable singularity formation result for expanding cosmological solutions to the
Einstein-scalar field and Einstein-stiff fluid systems
was proven in    \cite{RodnianskiSpeck}
for small perturbations of  the  spatially flat FLRW
solution with topology $R \times T^3$.
It was shown that  the  basic features of  the  perturbed solution in the
collapsing direction closely
resemble those of  the  FLRW solution,  
where its curvature blows up at a ``big bang''
singularity at $T^3$,
confirming Penrose's strong cosmic censorship hypothesis for
the "past-half" of near-FLRW solutions.   the  proof of linear stability
for  the  Einstein-scalar field system can be generalized  to control  the  non-linear terms.
Recently a proof of stable curvature blow up (without any symmetry assumptions)
towards  the  past singularity was given \cite{RS3}.

In  \cite{RodnianskiSpeck2} small perturbations of  the  family of
FLRW cosmological background
solutions to  the  Euler-Einstein system with a positive cosmological
constant were studied. It was shown that  the  background solutions, which describe an
initially uniform quiet fluid of positive energy density evolving
in a de Sitter spacetime undergoing accelerated expansion with
a linear equation of state,  are globally future asymptotically
stable under small irrotational perturbations.  In particular, it was
proven that  the  perturbed spacetimes, which have  the  topological
structure $R \times T^3$, are future causally geodesically
complete.

\newpage

\subsection{Spikes}

{\em{Spikes}}
are a generic feature of solutions of partial differential equations.
Therefore, it should be anticipated that spikes could be present in solutions of the field equations of GR. 
Indeed, in  the  oscillatory  approach to  an initial spacelike singularity, 
a spike could occur when a particular
point in the state space becomes stuck in a previous Kasner epoch while nearby points 
begin to evolve to a subsequent Kasner epoch. 
In this way, it is plausible that spatial derivatives might have a significant
effect at particular points in the early time cosmological dynamics.
Since spikes can be arbitrarily
narrow in the neighbourhood of the singularity, they are very difficult to simulate numerically.  
The mathematical analysis of spikes is also a challenge.
It is an open problem to prove  the  existence of spikes within GR and to determine their dynamical 
effect on generic inhomogeneous spacelike singularities.
Some mathematical justification for the existence of spikes
is known \cite{Ringstrom2004}.  
However, there has perhaps been greater success in obtaining
exact solutions with spike-like behaviour ~\cite{art:Lim2015}.

Spikes were first observed in  the  numerical  study of vacuum, orthogonally transitive,
spatially inhomogeneous $G_2$ cosmologies
~\cite{Berger,art:Limetal2009}. 
Studies of $G_2$ and subsequently  more general cosmological models therefore provide 
numerical evidence that the BKL dynamics generally occur except at very special isolated points 
(on one and two dimensional surfaces in three dimensional space) where spikes form \cite{art:Bergeretal2001}.
This then implies that  the asymptotic locality part of  the  BKL conjecture is violated.
Spikes also occur naturally in non-vacuum $G_2$ (and more general) spatially inhomogeneous
cosmological models. 
Due to gravitational instability, the spikes then leave
small imprints on the matter in  the  form of  residual perturbations. 
To date the main focus has been on investigating spikes
formed in  the  initial mixmaster regime and determining their imprint on matter and the subsequent
effect on structure formation \cite{art:ColeyLim2012}.

There are also  non-local recurring spikes associated with
cosmological singularities. However, it is believed that any spike formation
occurs only for a set of
time-lines of measure zero \cite{LARS99}. 
There are many other outstanding questions regarding recurring spike behaviour and generic spacelike singularities \cite{HeinzleUggla2012}.
In particular: How and where do spikes form? Do spikes experience  interference or annihilation?
Are there generic singularities that either do not have recurring spikes or
have a dense set of recurring spikes? 
And are there solutions that undergo an infinite number of recurring spike transitions?   
In addition, to date the dynamical behaviour of
solutions of  the  field equations 
in the asymptotic approach to a generic initial spacelike
singularity for both vacuum and simple
matter sources such as perfect fluids and scalar fields
\cite{art:LK63,WE} have been investigated. However,
it is also of interest, of course, to study  the  structural stability of generic spacelike singularities in the presence of matter such as 
electromagnetic fields and form fields (in which the  inhomogeneities  may influence the cosmological singularity quite differently).

%%\newpage

\subsection{Numerical relativity and relativistic astrophysics}

Numerical analysis has always played an important role 
in  any  problem within GR that concerns non-linear phenomena.
And recently there have been great advances in computational cosmology  \cite{Computational,Bentivegna}).
Indeed, numerics have been used to support many of  the  conjectures presented here and, in turn, have led to significant theoretical progress and to  the  formulation of new problems in mathematical GR.
We have discussed some of these problems earlier, including
the important question of critical phenomena in gravitational collapse,
which  was originally discovered
numerically \cite{Choptuik2015}.

Numerical analysis is also crucial for many problems in relativistic astrophysics. Indeed,
recently progress in numerical computations has facilitated the  investigation of  the  
inspiral and violent merger of a pair of compact objects (e.g., black 
holes or neutron stars), in which 
an enormous quantity of gravitational waves is emitted.
The determination of the  predictions of emitted waveforms  for binary  
systems for optimal
detection and parameter extraction is a primary goal.  
Such an analysis proved very important in the recent LIGO observations \cite{LIGO}.
More specifically,  the analysis of  the  gravitational-wave signals resulting from
black hole collisions give rise to very precise
theoretical predictions which were then used as 
template waveforms and cross-correlated with the data obtained from 
the gravitational-wave detectors.
In particular, analytic approximations
(such as post Newtonian 
expansions, black hole perturbation
theory and  the  effective one body approach) were used
to model the orbital dynamics and gravitational-wave emission, and
numerical relativity was then utilized to supplement these results  at late times close to coalescence
where such approximation schemes are no longer valid
\cite{Focusissue,Choptuik2015}.

To a large extent this problem has been solved in 
the case of a black-hole merger, although
the relatively simple properties of  the  two-body non-linear gravity waveforms \cite{YangPaschalidis}
have not been fully understood mathematically 
(and may involve new physics such as, for example, gravitational
turbulence \cite{refluis}
and is also related to  the  {\em{final state conjecture}}). 
The same is not true in  the  case of neutron stars. However, there the
difficulty is not with GR but with  the  matter;  the  binary
inspiral problem is a problem in mathematical physics
(utilizing appropriate approximation methods and numerical
techniques, and introducing  the  correct physics in  the  neutron star
case). Obtaining a  complete solution (spacetime) for the
two-body problem for
particles coming in from infinity in  the  past to a stationary
object to  the  future, is extremely difficult and beyond current analysis.
A number of open problems, particularly concerning the physical nature of the recent  
merger events and the 
gravitational waves detected by the LIGO-Virgo collaboration, 
have been discussed in \cite{Barack}.

%%%%%%%%%%%%%%%

\newpage

\section*{Acknowledgements}  
I would like to thank Piotr Bizon, Mihales Dafermos, Luis Lehner and Frans Pretorius 
for detailed comments on an earlier version of  the  manuscript, 
and Eric Poisson,  Claes Uggla and Clifford Will for useful remarks. 
Financial support was provided by  NSERC of Canada.

%%Malcolm, David G


\begin{thebibliography}{99}

\baselineskip 12pt


\bibitem{Yau1982}
S-T. Yau, ed.
``Seminar on differential geometry'', Annals of Math Studies
(Princeton University Press
Princeton, New Jersey, 1982).

\bibitem{Penrose1969} R.
Penrose, 
%%Gravitational collapse:  the  role of general relativity, 
Riv. Nuovo Cimento {\bf{1}} 252 (1969); R. Penrose, Gen. Rel. Grav. {\bf{34}} 1141 (2002);
R.Penrose, in ``Theoretical Principles in Astrophysics and Relativity'',
eds. N. R. Lebovitz, W. H. Reid and P. O. Vandervoort (Univ. of
Chicago Press, 1978)).



\bibitem{Penrose1973} R. Penrose, Ann. N.Y. Acad. {\bf{224}}  125 (1973).
%%%Naked singularities 


\bibitem{Bartnik} R. Bartnik,
``Some open problems in mathematical relativity'', Conference on Mathematical Relativity (Canberra, 1988),
ed. R. Bartnik, Proc. Centre Math. Anal. Austral. Nat. Univ., {\bf{19}}, Austral. Nat.
Univ., Canberra, 244–268 (1989). 
$https://projecteuclid.org/download/pdf_1/euclid.pcma/1416335857$


\bibitem{EJH}
J.-H. Eschenburg, 
%%The splitting theorem for spacetimes with strong energy condition, 
J. Diff. Geom. {\bf{27}} 477 (1988); see also
G.J. Galloway, 
%%Some connections between global hyperbolicity and geodesic completeness, 
J. Geom.
Phys. {\bf{6}} 127 (1989); R. P. A. C. Newman, 
%%A proof of  the  splitting conjecture of S.T. Yau, 
J. Diff. Geom.  {\bf{31}} 163 (1990). 

\bibitem{Galloway96}
 G. J. Galloway, ``Some rigidity results for spatially closed spacetimes'', pp. 21–34,  Mathematics of gravitation,
Part I (Polish Acad. Sci., Warsaw, 1997);
J.-H. Eschenburg and G. J. Galloway,
%% Lines in spacetimes, 
Comm. Math. Phys.
{\bf{148}} 209 (1992);
G. J. Galloway and C. Vega,
%% Achronal limits, lorentzian spheres, and splitting,
Ann. Henri Poincare {\bf{15}} 2241 (2014).
%%Hausdorff closed limits and rigidity in Lorentzian geometry
%% and [arXiv:1608.06353]. 

\bibitem{Focusissue} A. Le Tiec, B. Whiting and E. Poisson (eds.),
Focus issue: Approaches to  the  two-body problem;
http://www.iopscience.org/0264-9381/page/Focus Issues

\bibitem{LIGO}
B. P. Abbott {\em{et al.}} [LIGO/Virgo Collaboration], Phys. Rev. Lett. 
{\bf{116}}  061102 (2016)
[arXiv:1602.03837].

\bibitem{CGP}
 P. T.  Chrusciel, G. J.  Galloway and
D. Pollack, ``Mathematical general relativity:  A sampler'', Bull.  Amer.  Math.
Soc.  {\bf{47}}  567 (2010) [arXiv:1004.1016]. 

\bibitem{AretakisRodnianski} 
S. Aretakis and I Rodnianski,
``The Cauchy problem in general relativity'',  {\em{General relativity and gravitation}} (452–479, 
Cambridge Univ. Press, Cambridge, 2015): 
https://mathscinet.ams.org/mathscinet-getitem?mr=3644042. 


\bibitem{Coleyquestions} A. Coley, "Open problems in mathematical physics'',  Phys. Scr. {\bf{92}}  01 (2017).


\bibitem{Rendall2002} A. Rendall,  
%%Theorems on Existence and Global Dynamics for  the  Einstein field equations, 
Living Rev. Rel. {\bf{5}} 6 (2002)
[https://arxiv.org/abs/gr-qc/0203012].                                                                                 

\bibitem{LARS99}
L. Andersson, ``The global existence problem in general relativity,
The Einstein field equations and  the  large scale behaviour of gravitational fields'', pp. 71--120 (Birkhäuser, Basel,  2004) [arxiv/gr-qc/9911032].

\bibitem{Ionescu}
A. D. Ionescu and B. Pausader,
``On  the  global regularity for a Wave-Klein-Gordon coupled system''
[arXiv:1703.02846]; P. G. LeFloch and Y. Ma, 
%%The global nonlinear stability of Minkowski space for self-gravitating massive fields:  the  Wave-Klein-Gordon model
Comm. Math. Phys. {\bf{346}} 603 (2016).



\bibitem{Narita} M. Narita,
%%On  the  existence of global solutions for $T^{3}$-Gowdy spacetimes with stringy matter, 
Class.
Quant. Grav. {\bf{19}}  6279 (2002)
[arXiv:gr-qc/0210088].



\bibitem{CB69} Y. Choquet-Bruhat and R. Geroch, 
%%Global aspects of  the  Cauchy problemin general relativity, 
Comm. Math. Phys. {\bf{14}} 329 (1969).


\bibitem{CB80} Y.
Choquet-Bruhat and J. York, ``The Cauchy problem'', in 
``General relativity and gravitation'', vol. {\bf{1}} 99, ed. A. Held (Plenum, New
York, 1980).

\bibitem{Friedrich86}
H. Friedrich, 
%%On  the  existence of n-geodesically complete or future complete solutionsof Einstein’s field equations with smooth %%asymptotic structure, 
Comm. Math.
Phys. {\bf{107}} 587 (1986).

\bibitem{Dain}
S. Dain and H. Friedrich, 
%%Asymptotically flat initial data with prescribed regularity at infinity, 
Comm. Math. Phys. {\bf{222}} 569 (2001).

\bibitem{Kroon}
J. A. Valiente Kroon, 
%%A new class of obstructions to  the  smoothness of null infinity, 
Comm. Math. Phys. {\bf{244}}  133 (2004).

\bibitem{HansRingstrom17}
H. Ringstrom,
%% Origins and development of  the  Cauchy problem in general relativity
Class. Quant. Grav.   {\bf{32}}  124003 (2015) \&
``Linear systems of wave equations on cosmological backgrounds with convergent asymptotics''
[arXiv:1707.02803].

\bibitem{Ringstrom2015} 
H. Ringstrom,
``On proving future stability of cosmological solutions with accelerated expansion'' in  ``Surveys in 
differential geometry 2015: One hundred years of general relativity'', 
 Surv. Differ. Geom. {\bf{20}} 249 (Int. Press, Boston, MA, 2015).


\bibitem{ChristodoulouKlainerman90} D.
Christodoulou, and  S. Klainerman,
%%Asymptotic properties of linear field equations in Minkowski space, 
Commun. Pure Appl. Math. {\bf{43}}
137 (1990).

\bibitem{AnderssonMoncrief}
L. Andersson and V. Moncrief, ``Future complete vacuum spacetimes'', in  
``The Einstein equations and  the  large scale behaviour of gravitational fields'',
pp. 71--120 (Birkhäuser, Basel,  2004) [gr-qc/0303045]. 


\bibitem{Isenberg85}
J. Isenberg, 
%%Symmetries of cosmological Cauchy horizons with exceptional orbits, 
J. Math. Phys. {\bf{26}} 1024 (1985); V. Moncrief and J. Isenberg, 
%%Symmetries of cosmological Cauchy horizons,
Comm. Math. Phys. {\bf{89}} 387 (1983).


\bibitem{Chrusciel55}
P. T. Chrusciel and A. D. Rendall, 
%%Strong cosmic censorship in vacuum spacetimes with compact, locally homogeneous Cauchy surfaces, 
Ann. Physics {\bf{242}} 349  (1995).

\bibitem{Moncrief81}
V. Moncrief,
%% Global properties of Gowdy spacetimes with T3×R topology, 
Ann.
Phys. {\bf{132}} 87 (1981).

\bibitem{Chrusciel47}
P. T. Chrusciel, 
%%On spacetimes with U(1) × U(1) symmetric compact Cauchy surfaces, 
Ann. Phys. {\bf{202}} 100 (1990).



\bibitem{IsenbergMoncrief1982}
J. Isenberg and V. Moncrief, 
%%The existence of constant mean curvature foliations of Gowdy 3-torus spacetimes, 
Comm. Math. Phys. {\bf{86}} 485 (1982).

\bibitem{Chrusciel54}
P. T. Chrusciel, J. Isenberg and V. Moncrief, 
%%Strong cosmic censorship in polarised Gowdy spacetimes, 
Class. Quant. Grav. {\bf{7}} 1671 (1990).


\bibitem{Kichenassamy7}
S. Kichenassamy and A. D. Rendall, 
%%Analytic description of singularities in Gowdy spacetimes, 
Class. Quant. Grav. {\bf{15}}  1339 (1998).



\bibitem{Chrusciel49}
P. T. Chrusciel, ``On uniqueness in  the  large of solutions of Einstein’s equations; 
%%(“strong cosmic censorship”),
 Mathematical aspects of classical field theory" pp. 235–273 (Amer. Math. Soc., Providence, RI;  
Seattle, WA, 1991).

\bibitem{Chrusciel53}
P. T. Chrusciel and J. Isenberg, 
%%Nonisometric vacuum extensions of vacuum maximal globally hyperbolic spacetimes, 
Phys. Rev. D  {\bf{48}} 1616 (1993).

\bibitem{Choquet01}
Y. Choquet-Bruhat, 
%%Future global in time Einsteinian spacetimes with U(1) isometry group, 
Ann.
Henri Poincare {\bf{2}} 1007 (2001).

\bibitem{Choquet03}
Y. Choquet-Bruhat, ``Future complete Einsteinian space times with U(1) symmetry,
the unpolarized case'', 
The Einstein equations and  the  large scale behaviour of gravitational fields,
pp71--120, ed. L. Andersson (Birkhäuser, Basel, 2004) [[gr-qc/0305060]. 


\bibitem{Chrusciel52}
P. T. Chrusciel and G. J. Galloway, 
%%Horizons non-differentiable on a dense set, 
Comm. Math. Phys. {\bf{193}} 449 (1998).


\bibitem{Kunzinger}
M. Kunzinger, R. Steinbauer, and M. Stojkovic, 
Differential Geom. Appl. {\bf{34}}   14 (2014) \&
Class. Quant. Grav. {\bf{32  }}     155010 (2015)  [arXiv:1502.00287];
M. Kunzinger, R. Steinbauer,  M. Stojkovic and J. A. Vickers,
Class. Quant. Grav. {\bf{32 }}    075012 (2015)   [arXiv:gr-qc/1411.4689].



\bibitem{HughesKato}
T. J. R. Hughes, T. Kato and J. E. Marsden, 
%%Well-posed quasi-linear second-order hyperbolic systems with 
%%applications to nonlinear elastodynamics and general relativity, 
Arch. Rational Mech. Anal. {\bf{63}}  273 (1976).



\bibitem{KR}
S. Klainerman and I. Rodnianski, ``The causal structure of microlocalized Einstein
metrics'' (2001) [math.AP/0109174];
S. Klainerman and I. Rodnianski, 
%%Ricci defects of microlocalized Einstein metrics,
J. Hyperbolic Differ. Eqn. {\bf{1}} 85 (2004).

\bibitem{Tataru}
D. Tataru,
``Nonlinear wave equations'', Proceedings of  the  ICM, Beijing
{\bf{3}} (2003) [math.AP/0304397].


\bibitem{SenovillaGarfinkle}
J. M. M. Senovilla and D. Garfinkle, 
%%The 1965 Penrose singularity theorem 
%%Classical and Quantum Gravity for its "Milestones of General Relativity" focus issue
Class. Quant. Grav. {\bf{32}}  124008 (2015)
[arXiv:1410.5226]. 




\bibitem{Penrose65}
R. Penrose,
%%Gravitational Collapse and Spacetime Singularities. 
Phys. Rev. Lett. {\bf{14}} 57
(1965).

\bibitem{Penrose1979}
 R. Penrose, ``Singularities and time asymmetry'', in ``General Relativity: an Einstein Centenary
Survey'', eds.  S. W. Hawking and W. Israel (Cambridge University Press, Cambridge, 1979).




\bibitem{Hawking1966} S. W.
Hawking,  
%%The occurrence of singularities in Cosmology. I. 
Proc. Roy. Soc. London
{\bf{A294}}  511 (1966); {\em{ibid.}}, 
{\bf{A295}}  490 (1966);  {\em{ibid.}},   {\bf{ A300}} 
187 (1967).



\bibitem{PenroseHawking} R. Penrose and S. W. Hawking,
%%  the  Singularities Of Gravitational Collapse And Cosmology.
Proc. Roy. Soc. Lond. A {\bf{314}} 529 (1970).


\bibitem{HawkingEllis} S. W. Hawking and  G. F. R. Ellis , ``The large scale structure of spacetime'' (Cambridge Univ. Press, Cambridge, 1973).


\bibitem{Senovilla2012}
J. M. M. Senovilla, ``Singularity theorems in general relativity: achievements and open questions'',
Chapter 15 of Einstein and  the  Changing Worldviews of Physics, eds. C. Lehner, J. Renn and M.
Schemmel, Einstein Studies 12 (Birkhauser, 2012)



\bibitem{Christodoulou2009}
D. Christodoulou, ``The formation of black holes in general relativity'' (Monographs in Mathematics,
European Mathematical Soc. Publishing House, Helsinki, 2009).



\bibitem{Klainerman2014}
S. Klainerman, J. Luk and I. Rodnianski,
%% A fully anisotropic mechanism for formation of trapped surfaces in vacuum, 
Invent. Math. {\bf{198}} 1 (2014).


\bibitem{Klainerman2012}
S. Klainerman and I. Rodnianski, 
%%On  the  the formation of trapped surfaces, 
Acta Math. {\bf{208}} 211 (2012);
J. Luk and I. Rodnianski, ``Nonlinear interactions of impulsive gravitational waves for  the  vacuum Einstein
equations'',  Cambridge J. Math. [arXiv:1301.1072]; M. Dafermos, 
%%The formation of black holes in general relativity, 
Astrisque {\bf{123}} 352 (2013).

\bibitem{AnLuk} X. An and J. Luk, ``Trapped surfaces in vacuum arising dynamically from mild incoming radiation'' [arXiv:1409.6270]; X. An, ``Emergence of Apparent Horizon in Gravitational Collapse''[ arXiv:1703.00118].



\bibitem{AnderssonMetzger}
L. Andersson and J. Metzger, 
%%The area of horizons and  the  trapped region, 
Comm. Math. Phys., {\bf{290}}
941 (2009); M. Eichmair, 
%%The plateau problem for marginally trapped surfaces, 
J. Diff. Geom. {\bf{83}} 551
(2009); see also
S. Alexakis, 
%%``The Penrose inequality on perturbations of  the  Schwarzschild exterior'' 
Class. Quant. Grav. {\bf{33}}  115019  (2016)
[arXiv:1506.06400]; 
L. Andersson, M. Eichmair and J. Metzger, ``Jang’s equation and its applications to marginally trapped
surfaces'' [arXiv: 1006.4601].


\bibitem{Hamilton}
A. J. S. Hamilton and P. P. Avelino, 
%%The physics of  the  relativistic counter-streaming instability
%%that drives mass inflation inside black holes, 
Phys. Rept.
{\bf{495}} 1 (2010) [arXiv:0811.1926 gr-qc]; A. J. S. Hamilton,
``Inflation followed by BKL collapse inside accreting, rotating black holes''
[arXiv:1703.01921 gr-qc] 



\bibitem{MisnerTaub} C. Misner, ``Taub-NUT space as a counter-example to almost anything'', in
Relativity Theory and Astrophysics, ed. J. Ehlers (Amer. Math. Society,
1967).



\bibitem{Goncalves}
S. M. C. V. Goncalves,  
%%Naked singularities in Tolman-Bondi-de Sitter collapse
Phys. Rev. D {\bf{63}}  064017 (2001) [arXiv:gr-qc/0012032].


\bibitem{Christodoulou94} D.
Christodoulou,
%% Examples of naked singularity formation in  the  gravitational collapse of a scalar field, 
Ann. Math. {\bf{140}} 607 (1994).

\bibitem{Christodoulou99b} D.
Christodoulou,
%%  the  instability of naked singularities in  the  gravitational collapse of a scalar field, 
Ann. Math. {\bf{149}} 183 (1999).


\bibitem{Choptuik2015}
M. W. Choptuik, L. Lehner and F. Pretorius,
``Probing Strong Field Gravity Through Numerical Simulations'',
in ``General Relativity and Gravitation: A Centennial Perspective", eds. A. Ashtekar, B. 
Berger, J. Isenberg and M. A. H. MacCallum (Cambridge University Press, 2015) [arXiv:1502.06853]. 


\bibitem{Christod99}
 D. Christodoulou,  
 %%"On  the  global initial value problem and  the  issue of singularities",
Class. Quant.  Grav. {\bf{16}} A23 (1999).  



\bibitem{LukREF}
M. Dafermos and J. Luk,
``The interior of dynamical vacuum black holes I:   the  stability of  the  Kerr Cauchy horizon'' 
[arxiv/1710.01722].
%% authors also writing a survey for CQG on on strong cosmic censorship.
 

\bibitem{Ringstrom2010} H.
Ringstrom, 
%%Cosmic Censorship for Gowdy Spacetimes 
Living Rev. Rel.  {\bf{13}} 2 (2010).




\bibitem{Wald1998}
R. M. Wald, ``Gravitational collapse and cosmic censorship'', in ``Black Holes, Gravitational
Radiation and  the  Universe'', ed. by B.R. Iyer and B. Bhawal (Springer, Berlin, 1998) [arXiv: grqc/9710068].

\bibitem{Christod} D. Christodoulou, 
%%The problem of a self-gravitating scalar field, 
Comm. Math. Phys.
{\bf{105}} 337 (1986) \&
%%A mathematical theory of gravitational collapse, 
Comm. Math. Phys.
{\bf{109}} 613 (1987)


\bibitem{Isenberg92}
J. Isenberg, ``Progress on strong cosmic censorship, Mathematical aspects of classical
field theory'' pp. 403–418 (Amer. Math. Soc., Providence, RI; Seattle, WA, 1992).


\bibitem{Dias2015} O. J. C.
Dias, J. E. Santos  and B. Way, 
%%Black holes with a sine Killing vector field: black resonators, 
JHEP {\bf{12}} 171 (2015).

\bibitem{Horowitz2016} G. T. Horowitz, J. E. Santos and B. Way,
%%Evidence for an electrifying violation of cosmic censorship
Class. Quant. Grav. {\bf{33}} 195007 (2016). 

\bibitem{Bizon} P. Bizon and A. Rostworowski, 
%%On weakly turbulent instability of anti-de Sitter space, 
Phys.
Rev. Lett. {\bf{107}} 031102 (2011).



 
\bibitem{SchoenYau} R. Schoen and S. T. Yau, 
%%On  the  proof of  the  positive mass conjecture in general relativity, 
Comm. Math. Phys. {\bf{65}}  45 (1979) \&  
%%Proof of  the  positive mass theorem. II, 
Comm. Math. Phys. {\bf{79}} 
231 (1981).


\bibitem{Moncrief81E}
V. Moncrief and D. M. Eardley, 
%%The global existence problem and cosmic censorship in general relativity, 
Gen. Rel. Grav. {\bf{13}}  887 (1981).


\bibitem{Huisken}
G. Huisken and T. Ilmanen, 
%%The Riemannian Penrose inequality, Internat.
Math. Res. Notices  {\bf{20}}   1045 (1997) \&
%%and,  the  inverse mean curvature flow and  the  Riemannian Penrose inequality,
J. Diff. Geom. {\bf{59}} 353 (2001).


\bibitem{Moncrief96} A. D. Rendall, 
%%Constant mean curvature foliations in cosmological spacetimes, 
Helv. Phys.
Acta {\bf{69}} 490 (1996)
[gr-qc/9606049].



\bibitem{Bray} G. Huisken, ``An isoperimetric concept for mass and quasilocal mass'', (Oberwolfach Rep., no. 2, 87, 2006); H. L. Bray, J. Diff. Geom. {\bf{59}} 177 (2000).

%%``The Penrose inequality in general relativity and volume comparison theorems involving scalar
%%curvature'', ProQuest LLC, Ann Arbor, MI  (Ph.D. Thesis, Stanford University, 1997). 



\bibitem{NPC}
H. L. Bray and H. P. Roesch, ``Null Geometry and  the  Penrose Conjecture''
[arXiv:1708.00941];
H. Roesch,
``Proof of a null penrose conjecture using a new quasi-local mass''
[arXiv:1609.02875];
M. Mars and A. Soria,
%% On  the  penrose inequality along null hypersurfaces. 
Class. Quant. Grav.
{\bf{33}} 115019 (2016).



\bibitem{DainS}
S. Dain,
%%Geometric inequalities for black holes. 
Gen. Rel. Grav {\bf{46}}    1 (2014) [arXiv:1401.8166] \&
%%Geometric inequalities for axially symmetric black holes.
Class. Quant. Grav. {\bf{ 29}}   073001  (2012);
M. Mars and A. Soria, 
%%“On  the  Penrose Inequality Along Null Hypersurfaces”, 
Class. Quant. Grav. {\bf{ 33  }}   115019 (2016) [arXiv:1511.06242/gr-qc].




\bibitem{Tod}
P. Tod, 
%%Some examples of Penrose's quasi-local mass construction, 
Proc.
Roy. Soc. Lond. A {\bf{388}} 457 (1983).



\bibitem{Gibbons} 
G. Gibbons, ``The isoperimetric and Bogomolny inequalities for black holes'',
in ``Global Riemannian Geometry'', ed. T. Willmore and N. Hitchin (Ellis
Harwood Ltd., Chichester, 1984).


\bibitem{Thorne1972} K. S.
Thorne, ``Nonspherical Gravitational Collapse - A Short Review'', in ``Magic without Magic:
John Archibald Wheeler'',  p. 231 ed. J. Klauder (Freeman, S. Francisco, 1972).


\bibitem{Senovilla2008} J. M. M.
Senovilla, 
%% A reformulation of  the  Hoop Conjecture 
Europhys. Lett. {\bf{81 }} 
20004 (2008).




\bibitem{MWChoptuik} M. W. Choptuik, Phys. Rev. Lett. {\bf{70}} 9 (1993).


\bibitem{Reiter} M. Reiterer and E. Trubowitz,
``Choptuik's critical spacetime exists''
[arXiv:1203.3766].
%%  -- no ref; a new version of this paper is now available.


\bibitem{BartnikMckinnon}
R. Bartnik and J. Mckinnon, 
%%Particle-like solutions of  the  Einstein Yang-Mills equations, 
Phys. Rev. Lett. {\bf{61}} 141 (1988).


\bibitem{Eardley} D.
Eardley and V. Moncrief,  
%%The global existence of Yang– Mills–Higgs fields in 4-dimensional Minkowski space. I. 
%%Local existence and smoothness properties, 
Comm. Math. Phys. {\bf{83}} 171 (1982)
%%; II. Completion of proof, 
\& Comm. Math. Phys. {\bf{83}} 
193 (1982).

\bibitem{Klainerman95} S.
Klainerman and M. Machedon, 
%%Finite energy solutions of  the  Yang-Mills equations, 
Ann. Math. {\bf{142}} 39 (1995).

\bibitem{Chrusciel56}
P. T. Chrusciel and J. Shatah, 
%%Global existence of solutions of  the  Yang-
%%Mills equations on globally hyperbolic four-dimensional Lorentzian manifolds, 
Asian
J. Math. {\bf{1}} 530 (1997).

\bibitem{Friedrich91}
H. Friedrich,
%%On  the  global existence and  the  asymptotic behaviour of solutions to the
%%Einstein-Maxwell-Yang-Mills equations, 
J. Diff. Geom. {\bf{34}} 275 (1991).



\bibitem{Olinyk}
R. A. Bartnik, M. Fisher and T. A. Olinyk,
%%Static Spherically Symmetric Solutions of  the  SO(5) Einstein Yang-Mills Equations,
J. Math. Phys. {\bf{51}} 032504 (2010) 
[arXiv:0907.3975].






\bibitem{Bizonwavemaps} P.
Bizon, 
%%Equivariant self-similar wave maps from Minkowski spacetime into 3-sphere, 
Comm. Math. Phys. {\bf{215}} 45 (2000); P.
Bizon, T. Chmaj and  Z. Tabor, 
%%Formation of singularities for equivariant 2+1 dimensional wave maps into 2-sphere, 
Nonlinearity {\bf{14}} 1041 (2001).



\bibitem{AnderssonGudapatiSzeftel}
L. Andersson, N. Gudapati and J. Szeftel, 
``Global Regularity for  the  2+1 Dimensional Equivariant Einstein-Wave Map System'' 
[arXiv:1501.00616].


\bibitem{SterbenzTataru}
J. Sterbenz and D. Tataru,
%% Regularity of Wave-Maps in Dimension 2 + 1 
Comm. Math. Phys. {\bf{298}}    231 (2009) 
[arXiv:0907.3148]; J. Krieger and W. Schlag, 
``Large global solutions for energy supercritical nonlinear wave equations on $R^{3+1}$''
[arXiv:1403.2913]. 


\bibitem{BizonBiernat}
P. Bizon and P. Biernat, Comm. Math. Phys.  s00220-015-2404-y (2015); see also 
P. Bizon, Acta Physica Polonica  B {\bf{33}} 1893 (2002).

\bibitem{Donninger} R. Donninger, Commun. Pure Appl. Math. {\bf{64}} 1095 (2011) \&
Math Z. {\bf{278}} 1005 (2014).

\bibitem{Chmaj}
M. Choptuik, T. Chmaj and P. Bizon,
Phys. Rev. Lett. {\bf{77}} 424 (1996) [arXiv:gr-qc/9603051].
%%Critical behaviour in Gravitational Collapse of a YM Field 




\bibitem{Andreasson}
H. Andreasson, 
%%``The Einstein-Vlasov System/Kinetic Theory'', 
 Living Rev. Rel. {\bf{14 }}  4 (2011) 
[arXiv:1106.1367].



\bibitem{Christodoulou93}
D. Christodoulou and S. Klainerman, ``The global nonlinear stability of  the  Minkowski space''
(Princeton mathematical series, 41, Princeton University Press, 1993).


\bibitem{Bieri}
H. Lindblad and I. Rodnianski, 
%%Global existence for  the  Einstein vacuum equations in wave coordinates, 
Comm.
Math. Phys. {\bf{256}} 43 (2005); L. Bieri and N. Zipser,
``Extensions of  the  stability theorem of  the  Minkowski space in general relativity'', AMS/IP Studies
in Advanced Mathematics, 45  (American Mathematical Society, Providence, RI; International Press, Cambridge,
MA, 2009).

\bibitem{EinsteinVlasov}
D. Fajman, J. Joudioux and J. Smulevici,
``The Stability of  the  Minkowski space for  the  Einstein-Vlasov system''
[arXiv:1707.06141];
H. Lindblad and M. Taylor, Commun. Part. Diff. Eq.
[arXiv:1707.06079]; 
%%Global stability of Minkowski space for  the  Einstein--Vlasov system in  the  harmonic gauge
M. Taylor, Ann. PDE {\bf{3}} 9 (2017) [arXiv:1602.02611].
%%The global nonlinear stability of Minkowski space for  the  massless Einstein--Vlasov system



\bibitem{Bunting}
G. Bunting and A. K. M. Masood-ul-Alam, 
%%Non-existence of multiple black holes in asymptotically Euclidean static vacuum spacetime, 
Gen. Rel. Grav. {\bf{19}}
147 (1987).

\bibitem{Israel}
W. Israel,
%% Event horizons in static electrovac spacetimes, 
Comm. Math.
Phys. {\bf{8}}  245 (1968).

\bibitem{Kerr}
R. P. Kerr, Phys. Rev. Lett. {\bf{11}} 237 (1963).

\bibitem{Carter}
 B. Carter, in Black Holes, 1972 Les Houches Lectures, eds. B. DeWitt
and C. DeWitt (Gordon and Breach, NY, 1973); B. Carter, 
%%Bunting identity and Mazur identity for nonlinear elliptic systems,
%%including  the  black hole equilibrium system, 
Comm. Math. Phys. {\bf{99}}
563 (1985).

\bibitem{Robinson}
D. C. Robinson, 
%%Uniqueness of  the  Kerr black hole, 
Phys. Rev. Lett. {\bf{34}} 905 (1975).



\bibitem{Newman65} E. Newman, J. Math. Phys. {\bf{6}} 918 (1965).


\bibitem{Mazur}
P. Mazur,
%% Proof of uniqueness of  the  Kerr-Newman black hole solution,
J. Phys. A {\bf{15}} 3173 (1982).




\bibitem{Heusler}  M. Heusler,
Living Rev. Rel. {\bf{1}} 6 (1998);
http://www.livingreviews.org/Articles/Volume1/1998-6heusler.

\bibitem{Chand}
S. Chandrasekhar, ``Mathematical Theory of Black Holes'' (Oxford University Press, 1983).

\bibitem{Holzegel1016}
G. Holzegel,
%%conservation laws
Class. Quant. Grav. {\bf{33}}  205001 (2016).


\bibitem{DHR}
M. Dafermos, G. Holzegel and I. Rodnianski, ``The linear stability of  the  Schwarzschild solution to gravitational perturbations'',
146 pages (2016) [arXiv:1601.06467].


\bibitem{KlainermanSzeftel}
 S. Klainerman and J. Szeftel, 
``Global Nonlinear Stability of Schwarzschild Spacetime under Polarized Perturbations'',
425 pages [arXiv:1711.07597].

\bibitem{Shlapentokh}
M. Dafermos, I. Rodnianski and Y. Shlapentokh-Rothman, ``Decay for solutions of
the wave equation on Kerr exterior spacetimes:  the  full subextremal case''
[arXiv:1402.7034].

\bibitem{lecturenotes}
M. Dafermos and I. Rodnianski,
``Lectures on black holes and linear
waves'', Clay Mathematics
Proceedings {\bf{17}} 97 (2008) [arXiv:0811.0354].

\bibitem{recentpaper}
M. Dafermos, G. Holzegel and I. Rodnianski,
 ``Boundedness and decay for  the  Teukolsky equation on Kerr spacetimes I'' 
[arxiv/1711.07944]. 




\bibitem{Zilho}
M. Zilhao, V. Cardoso, C. Herdeiro, L. Lehner and U. Sperhake, 
Phys. Rev. D {\bf{90}} 124088  (2014) [arXiv:1410.0694].
%%Testing  the  nonlinear stability of Kerr-Newman black holes 


\bibitem{Friedman}
J. L. Friedman, Comm.  Math. Phys. 
{\bf{63}} 243 (1978)
[arxiv/1608.02035].
%%% Ergosphere instability.


\bibitem{Moschidis16} G. Moschidis, 
``A proof of Friedman's ergosphere instability for scalar waves''
[arxiv/1608.02035].
%%,  -- no ref

\bibitem{Aretakis}
S. Aretakis,
``Horizon Instability of Extremal Black Holes''
[arxiv/1206.6598].
%%  -- no ref.

\bibitem{Hintz}
P. Hintz and A. Vasy, ``The global non-linear stability of  the  Kerr-de Sitter family of black
holes'' [arXiv:1606.04014];
V. Schlue, 
%%Global results for linear waves on expanding Kerr and Schwarzschild-de Sitter cosmologies, 
Comm. Math. Phys. {\bf{334}} 977 (2015).


\bibitem{Holzegel} G. Holzegel and J. Smulevici,
%% Quasimodes and a lower bound on  the  uniform energy decay rate for Kerr-AdS spacetimes, 
Anal. PDE {\bf{7}} 1057 (2014)
[arXiv:1303.5944].

\bibitem{DafermosHolzegelRodnianski} 
M. Dafermos, G. Holzegel and I. Rodnianski,
``A scattering theory construction of dynamical vacuum black holes'',
to appear in J. Diff. Geom. (2013) [arXiv:1306.5364].

\bibitem{YangPaschalidis}
H. Yang, V. Paschalidis, K. Yagi, L. Lehner, F. Pretorius and N. Yunes, 
Phys. Rev. D {\bf{97}} 024049 (2018) [arXiv:1707.00207].
%%Gravitational wave spectroscopy of binary neutron star merger remnants with mode stacking 


\bibitem{Ehlers} J.
Ehlers, ``The Newtonian limit of general relativity'', in ``Classical mechanics and relativity: relationship and consistency'', ed. G. Ferrarese, (Bibliopolis,
Naples, 1991); see also
J. Ehlers and  T. Buchert,
%%Newtonian Cosmology in Lagrangian Formulation: Foundations and Perturbation Theory. 
Gen. Rel. Grav. {\bf{29}} 733 (1997); T. A. Oliynyk, 
%%The Newtonian limit for perfect fluids ,
Comm. Math. Phys. {\bf{276}} 131 (2007) 
[arXiv:astro-ph/9510056].

\bibitem{JEhlers} J. Ehlers, 
``Folklore in relativity and what is really known'', in General Relativity and Gravitation'' {\bf{11}}, 
ed. M.A. H. MacCallum (Springer, Stockholm 1986).

\bibitem{BartnikPollack}
R. Bartnik,
%% Remarks on cosmological spacetimes and constant mean curvature surfaces. 
Comm. Math. Phys.
{\bf{117}} 615 (1988);
P. T. Chru´sciel, J. Isenberg and D. Pollack,
%% Initial data engineering. 
Comm. Math. Phys. {\bf{257}} 29 (2005).


\bibitem{DiltsHolst}
J. Dilts and M. Holst, ``When Do Spacetimes Have Constant Mean Curvature Slices?''
[arXiv:1710.03209].



\bibitem{CHP}
A. Coley, S. Hervik and N. Pelavas, Class. Quant. Grav. {\bf 26} 025013 (2009)
[arXiv:0904.4877].

\bibitem{CHP2} A. Coley, S. Hervik and N. Pelavas, 
%%Lorentzian  manifolds and scalar curvature invariants, 
Class. Quant. Grav. 
{\bf{27}} 102001 (2010)
[arXiv1003.2373]; see also 
A. Coley and S. Hervik, 
%%Discriminating  the   Weyl type in higher dimensions using scalar curvature invariants,  
Gen. Rel. Grav.
{\bf{43}} 2199 (2011).

\bibitem{krameretal}
H. Stephani, D. Kramer, M. MacCallum, C. Hoenselaers and E. Herlt, ``Exact Solutions of Einstein’s
Field Equations'' (Cambridge University Press, Cambridge, second ed., 2003).


\bibitem{ColeyMcNutt} A. Coley, D. D. McNutt, and A. Shoom, 
%%Geometric horizons, 
Phys. Lett. B {\bf{771}} 131
(2017); A. Coley and D. D. McNutt,
%%``Identification of black hole horizons using discriminating scalar curvature i 2018,
Class. Quant. Grav. {\bf{35}}  025013 (2018)  [arXiv:1710.08773].


\bibitem{BGYone}
L. Bieri, D. Garfinkle and N. Yunes, AMS Notices {\bf{64}} 07 (2017)
[arXiv:1710.03272];
%%Gravitational Waves and Their Mathematics
L. Bieri, D. Garfinkle and S.-T. Yau,
in  ``The Centenary of General Relativity'', volume 20 of ``Surveys in Differential Geometry'' 
[arXiv:1505.05213]. 
%%Gravitational Waves and Their Memory in General Relativity


\bibitem{Christodoulou1991}
 D. Christodoulou, Phys. Rev. Letts. {\bf{67}} 1486 (1991).
%% Nonlinear Nature of Gravitation and Gravitational-Wave Experiments. 

\bibitem{BGYtwo}
 L. Bieri, D. Garfinkle and N. Yunes, ``Gravitational wave memory in de Sitter spacetime''
[arXiv:1706.02009]; 
L. Bieri, D. Garfinkle and S.-T. Yau,
 Phys. Rev. D {\bf{94}} 064040 (2016)  [arXiv:1509.01296].
%%

\bibitem{Ishak}
M. Ishak, ``Testing general relativity in cosmology''
[arXiv:1806.10122].



\bibitem{Goode} S. W.
Goode and  J. Wainwright, 
%%Isotropic singularities in cosmological models, 
Class. Quant. Grav. {\bf{2}} 99 (1985);
S. W. Goode, A. A. Coley and J. Wainwright,
%%The isotropic singularity in cosmology, 
Class. Quant. Grav. {\bf{9}} 445 (1992)
[arXiv:0810.3744]; C. M.
Claudel and K. P. Newman, 
%%The Cauchy problem for quasi-linear hyperbolic evolution problems with a singularity in  the  time, 
Proc. R.
Soc. London, Ser. A {\bf{454}} 3 (1998).
 

\bibitem{Middleton}
J, Middleton and J. D. Barrow,
%%The Stability of an Isotropic Cosmological Singularity in Higher-Order Gravity,
Phys. Rev. D {\bf{77}} 10352 (2008) [arXiv:0801.4090];
I. V. Kirnos, A. N. Makarenko, S. A. Pavluchenko and A. V. Toporensky,
%%The nature of singularity in multidimensional anisotropic Gauss-Bonnet cosmology with a perfect fluid,
Gen. Rel. Grav. {\bf{42}} 2633 (2010) [arXiv:gr-qc/0906.0140];
J. D. Barrow and S. Hervik,
%%Simple Types of Anisotropic Inflation,
Phys. Rev. D {\bf{81}} 023513 (2010) 
[arXiv:0911.3805] 

\bibitem{Hawking}
S. W. Hawking, Nature {\bf{248}}, 30 (1974); S. W. Hawking, Phys. Rev. {\bf{D14}}, 2460 (1976);
S. Hawking, 
%%Particle Creation by Black Holes, 
Comm. Math. Phys. {\bf{43}} 199 (1975).

\bibitem{Barcelo}
C Barcelo,
R Carballo-Rubioy and
L. J. Garayz, JHEP {\bf{05}} 054  (2017)
[arXiv:1701.09132].

\bibitem{mald}  J. M. Maldacena,  
Int. J. Theor. Phys. {\bf{38}} 1113 
(1999); J. M. Maldacena, Adv. Theor. Math. Phys. {\bf{2}} 231 (1998).
%%  the  large N limit of superconformal field theories and supergravity


\bibitem{Klebanov}
I. Klebanov and J. Maldacena,  
%%Solving Quantum Field Theories via Curved Spacetimes, 
Physics Today {\bf{62}} 
28 (2009).




\bibitem{HolzegelShao}
G. Holzegel and A. Shao, ``Unique continuation from infinity in asympotically Anti-de Sitter spacetimes 
II: Non-static boundaries'' [arXiv:1608.07521/gr-qc].




\bibitem{Friedrich}
 H. Friedrich, 
 %%Einstein equations and conformal structure: existence of anti-de Sitter- type spacetimes, 
 J. Geom. Phys. {\bf{17}} 125 (1995)



\bibitem{PiotrBizon} P. Bizon, 
%%Is AdS stable?  (invited review for GRG based on plenary lecture at GR20), 
Gen. Rel. Grav. {\bf{46}} 1724 (2014) [arXiv:1312.5544].


\bibitem{Bizon2015} P.
Bizon, M. Maliborski and  A. Rostworowski, Phys. Rev. Lett. {\bf{115}} 081103 (2015).


\bibitem{Dafermos}
 M. Dafermos and  G. Holzegel 2006 Seminar at DAMTP (University of Cambridge) available at:
https://dpmms.cam.ac.uk/~md384/ADSinstability.pdf


\bibitem{Anderson} M. T. Anderson, 
%%On  the  uniqueness and global dynamics of AdS spacetimes, 
Class.
Quant. Grav. {\bf{23}} 6935 (2006).


\bibitem{MalRost}
M. Maliborski and
A. Rostworowski, Phys.  Rev. Lett.  {\bf{111}} 051102 (2013)
[arXiv:1303.3186]. 
%%Time-periodic solutions in Einstein AdS - massless scalar field system:  


\bibitem{DiasSantos}
O. J. C. Dias, G. T. Horowitz and J. E. Santos, 
%%Gravitational Turbulent Instability of Anti-de Sitter Space, 
Class. Quant. Grav. {\bf{29}} 194002  (2012)  [arXiv:1109.1825];
O. J. C. Dias, and J. E. Santos,
%%AdS nonlinear instability: moving beyond spherical symmetry,
%%Class. Quant. Grav. {\bf{33}} 23LT01 (2016) \& 
``AdS nonlinear instability: breaking 
spherical and axial symmetries'' [arXiv:1705.03065]; 
A. Rostworowski, 
%%Comment on "AdS nonlinear instability: moving beyond spherical symmetry" 
Class. Quant. Grav. {\bf{33}} 23LT01 (2016)] [arXiv:1612.00042].


\bibitem{DiasGreen}
O. J. C. Dias, G. T. Horowitz, D. Marolf and  J. E. Santos, Class. Quant. Grav. {\bf{29}}  235019  (2012);
%%On  the  Nonlinear Stability of Asymptotically Anti-de Sitter Solutions
S. R. Green, A. Maillard, L. Lehner and S. L. Liebling,  
Phys. Rev. D {\bf{92}} 084001 (2015) [arXiv:1507.08261].


\bibitem{Martinon} G. Martinon,
``The instability of anti-de Sitter space-time''
 [arXiv:1708.05600].



\bibitem{Moschidis}
G. Moschidis, ``The Einstein null dust system in spherical symmetry with an inner mirror: structure of  the  maximal
development and Cauchy stability'' [arXiv:1704.08685] \&
``A proof of  the  instability of AdS for  the  Einstein null dust system with an inner mirror''
[arXiv:1704.08681].





\bibitem{Jalmuzna} J. Jalmuzna, A. Rostworowski and P. Bizon, 
%%AdS collapse of a scalar field in higher dimensions, 
Phys. Rev. D {\bf{84}} 085021 (2011).



\bibitem{string}
M Green,  J Schwarz and  E Witten, ``Superstring Theory'' (Cambridge: Cambridge
University Press, 1988);  J. Polchinski, ``String Theory'' (Cambridge: Cambridge University Press 2005)



\bibitem{EmparanReall} 
R. Emparan and H. S. Reall,
%% Black Holes in Higher Dimensions,
Living Rev. Rel. {\bf{11}} 6 (2008) [arXiv:0801.3471].




\bibitem{higherdimensions}
A. Coley, R. Milson, V. Pravda and A. Pravdova, 
%%Classification of  the  Weyl tensor in higher  dimensions, 
Class. Quant. Grav. {\bf{21}} L35  (2004)  [gr-qc/0401008]; 
A. Coley, 
%%Classification of  the  Weyl tensor in higher dimensions and applications, 
Class. Quant. Grav. {\bf{25}}  033001 (2008) [arXiv:0710.1598].
                




\bibitem{Galloway}
G. J. Galloway and J. M. M. Senovilla, 
%% Singularity theorems based on trapped submanifolds of arbitrary co-dimension, 
Class. Quant. Grav. {\bf{27}}  152002 (2010).


\bibitem{SchoenYau2017}
R. Schoen and S.-T. Yau. ``Positive Scalar Curvature and Minimal Hypersurface Singularities''
[arXiv:1704.05490].



\bibitem{LehnerPretorius2010}
L. Lehner and F. Pretorius,  
%% Black strings, low viscosity fluids, and violation of cosmic censorship
Phys. Rev. Lett. {\bf{105}} 101102
(2010).


\bibitem{GregoryLaflamme}
R. Gregory and R. Laflamme,  
%%Black strings and p-branes are unstable, 
Phys. Rev. Lett. {\bf{70}} 2837
(1993).



\bibitem{EmparanReall2002}
R. Emparan and H. S. Reall, Phys. Rev. Lett. {\bf{88}} 101101
(2002).



\bibitem{BRI}
J. E. Santos and B. Way, Phys. Rev. Lett. {\bf{114}}, 221101 (2015);
K. Tanabe, JHEP {\bf{02}} 151 (2016);
P. Figueras, M. Kunesch, and S. Tunyasuvunakool, Phys. Rev.
Lett. {\bf{116}} 071102 (2016).

\bibitem{MP}
R. C. Myers and M. J. Perry, Ann. Phys. {\bf{172}} 304 (1986).

\bibitem{EmparanMyers}
R. Emparan and R. C. Myers, JHEP {\bf{09}} 025 (2003).



\bibitem{Figueras} O. J. C. Dias, P. Figueras, R. Monteiro, J. E. Santos, and
R. Emparan, Phys. Rev. D   {\bf{80}} 111701 (2009);
P. Figueras, M. Kunesch, L. Lehner, and S. Tunyasuvunakool, 
%%End Point of  the  Ultraspinning Instability and Violation of Cosmic Censorship,
Phys. Rev. Letts. {\bf{118}} 151103 (2017).


\bibitem{Henneaux}
M. Henneaux,
%%Kac-Moody algebras and  the  structure of cosmological singularities: a new light on  the  Belinskii-
Khalatnikov-Lifshitz analysis 
in {\em{Quantum Mechanics of Fundamental Systems:  the  Quest for Beauty and Simplicity - Claudio Bunster 
Festsschrift}} [arXiv:0806.4670]. 

\bibitem{infl}  W.  E.  East, M.  Kleban, A.  Linde and L.  Senatore, 
JCAP {\bf{09}} 010 (2016) [arXiv:1511.05143]; 
J. Braden, M. C. Johnson, H. V. Peiris and A. Aguirre,
Phys. Rev. D {\bf{96}} 023541 (2017) [arXiv:1604.04001].


\bibitem{Brandenberger}
R. Brandenberger and P. Peter,  Found. Phys. {\bf{47}} 797 (2017) [arXiv:1603.05834 hep-th].

\bibitem{EKp}
D. Garfinkle, W. C. Lim, F. Pretorius and P. J. Steinhardt,  Phys. Rev. D {\bf{78}} 083537 (2008);
B. Xue, D. Garfinkle, F. Pretorius and P. J. Steinhardt, Phys. Rev. D {\bf{88}} 083509 (2013).

\bibitem{AL42} R. E. Allen and
S. Lidstrom,``Life,  the  universe, and everything: 42 fundamental
questions'',  Phys. Scr. {\bf{92}}  012501 (2017)
[Focus Issue on 21st Century Frontiers]. 


\bibitem{Freese}
K.~Freese,
%%Observational Status of Dark Matter in  the  Universe
%%Physica Scripta to appear (2017); see also 
``Status of Dark Matter in  the  Universe'' [arXiv:1701.01840].
%%`Status of Dark Matter in  the  Universe,'' to be published in Conference Proceedings of  the  
%%Fourteenth Marcel Grossmann Meeting on General Relativity, Rome, July 2015



\bibitem{Witten2001}
E. Witten, ``The cosmological constant from the
viewpoint of string theory'', in 
Sources and Detection of Dark Matter and Dark Energy
in  the  Universe, ed. D. B. Cline pages 27--36 (Springer, Berlin, Heidelberg,
2001).


\bibitem{Steinhardt}
P. Steinhardt and N. Turok,   
%% "Why  the  Cosmological constant is so small and positive".  
Science {\bf{312}}  1180  (2006) [arXiv:astro-ph/0605173].



\bibitem{Weinberg1989}
S. Weinberg, 
%%The cosmological constant problem, 
Rev. Mod. Phys. {\bf{61 }} 1 (1989).


\bibitem{Padilla}
A. Padilla, ``Lectures on  the  Cosmological Constant
Problem'' [arXiv:1502.05296].


\bibitem{Weinberg1987} S. Weinberg, 
%%Anthropic bound on  the  cosmological constant 
Phys. Rev. Lett. {\bf{59}} 2607 (1987).

\bibitem{Riess}
A. G. Riess {\em{et al.}},  
%%Observational evidence from supernovae for an accelerating Universe and a cosmological constant, 
Astron. J. {\bf{116}} 1009 (1998);
S. Perlmutter {\em{et al.}}, 
 %%Measurement of Ω and Λ from 42 high-redshift supernovae 
Astrophys.
J. {\bf{517}} 565 (1999).



\bibitem{tension} 
T. Buchert, A. A. Coley, H. Kleinert, B. F. Roukema and D. L. Wiltshire, Int. J. Mod. Phys.
D {\bf{25}} 1630007  (2016) [arXiv:1512.03313].


\bibitem{web}
J.~Einasto,
``Yakov Zeldovich and  the  Cosmic Web Paradigm'',
in Proc. IAU Symp. {\bf 308}, eds.
R.~van de Weygaert, S.~Shandarin, E.~Saar, J.~Einasto 
(Cambridge Univ. Press, 2017)
[arXiv:1410.6932];
%%CITATION = ARXIV:1410.6932;%%
F.~Hoyle and M. S.~Vogeley,
%{\em``Voids in  the  Point Source Catalogue Survey and  the  Updated Zwicky
%Catalog''},
Astrophys. J. {\bf 566}  641 (2002) [arXiv:astro-ph/0109357];
%%CITATION = ASTRO-PH 0109357;%%
%{\em``Voids in  the  2dF Galaxy Redshift Survey''}, \&
Astrophys. J. {\bf 607}  751 (2004) [arXiv:astro-ph/0312533];
%%CITATION = ASTRO-PH 0312533;%%
D. C.~Pan, M. S.~Vogeley, F.~Hoyle, Y. Y.~Choi, and C.~Park,
%{\em``Cosmic voids in Sloan Digital Sky Survey Data Release 7,''}
Mon. Not. R. Astron. Soc. {\bf 421}  926 (2012) [arXiv:1103.4156].
%\href{http://arxiv.org/abs/arXiv:1103.4156}
%%CITATION = ARXIV:1103.4156;

\bibitem{Av} 
%%R. van den Hoogen {\em{et al.}}, preprint.
J. Brannlund, R. van den Hoogen and A. Coley,
``Averaging geometrical objects on a differentiable manifold'' [arXiv:1003.2014];
R. van den Hoogen, J. Math. Phys. {\bf{58}} 122501
(2017); A. A. Coley,
%%Averaging in cosmological models using scalars
Class. Quant. Grav. {\bf 27} 245017 (2010) [arXiv:0908.4281]; see also
T.  Buchert {\em{et al.}},  Class.  Quant. Grav.  {\bf{32}}  215021 (2015)
[arXiv:1505.07800].

\bibitem{Averaging} R. M. ~Zalaletdinov,
%{\em``Averaging out  the  Einstein equations and macroscopic spacetime
%geometry''},
Gen. Rel. Grav. {\bf 24} 1015 (1992) \&
%{\em``Towards a theory of macroscopic gravity''},
Gen. Rel. Grav. {\bf 25}  673  (1993) [arXiv:gr-qc/9703016];
M. Mars and R. M. Zalaletdinov, J. Math. Phys. {\bf 38}  4741 (1997);
A. A.~Coley, N. ~Pelavas and R. M.~Zalaletdinov,
Phys. Rev. Letts. {\bf 95} 151102  (2005) [arXiv:gr-qc/0504115].





\bibitem{LAndersson14}
L. Andersson,
``Cosmological Models and Stability '', in 
 ``General Relativity, Cosmology and Astrophysics, Fundamental Theories of Physics'',  {\bf{177}} p. 277 
(Springer International Publishing Switzerland, 2014; ISBN 978-3-319-06348-5). 



\bibitem{BarrowTipler}
J. D. Barrow, G. J. Galloway and F. J. Tipler, 
%%The closed-universe recollapse conjecture, 
Mon. Not. R. Astron. Soc. {\bf{223}}  835 (1986).

\bibitem{LinWald}
X. Lin and R. M. Wald, 
%%Proof of  the  closed-universe-recollapse conjecture for diagonal Bianchi type-IX cosmologies, 
Phys. Rev. D {\bf{40}} 3280 (1989) \& 
{\bf{41}} 2444 (1990). 


\bibitem{Friedrich1986} H.
Friedrich,
%% Existence and structure of past asymptotically simple solutions
%%of Einstein’s field equations with positive cosmological constant,
J. Geom. Phys. {\bf{3}} 101 (1986).


\bibitem{Wald83} R. Wald,
 %%Asymptotic behaviour of homogeneous cosmological models with cosmological constant, 
 Phys. Rev. D {\bf{28}} 2118 (1983).

\bibitem{Rendall95}  A. D. Rendall, 
%%Global properties of locally homogeneous cosmological models with matter., 
Math. Proc. Camb. Phil. Soc. {\bf{118}}  511
(1995).

\bibitem{Coleybook}
A. A. Coley, 
``Dynamical systems and cosmology'' (Kluwer Academic,
Dordrecht: ISBN 1-4020-1403-1, 2003). 

\bibitem{exppot} J. M.
Heinzle and A. D. Rendall, 
%%Power-law Inflation in Spacetimes without Symmetry. 
Comm.
Math. Phys. {\bf{269}} 1 (2007); H.
Ringstrom,
%%Power law inflation. 
Comm. Math. Phys. {\bf{290}} 155 (2009).


\bibitem{Jensen} L. G.
Jensen and  J. A. Stein-Schabes,
%%Is inflation natural?, 
Phys. Rev.
D {\bf{35}} 1146 (1987).




\bibitem{art:LK63}
E.~M. Lifshitz and I.~M. Khalatnikov,
Adv. Phys. {\bf 12} 185 (1963);
V.~A. Belinskii, I.~M. Khalatnikov, and E.~M. Lifschitz,
Adv. Phys. {\bf 19}, 525 (1970); {\em{ibid.}}
{\bf 31} 639 (1982);
V.~A. Belinskii and I.~M. Khalatnikov, 
Soviet Scientific Review Section A: Physics Reviews {\bf 3} 555 (1981).

\bibitem{Berger}
B. K. Berger and V. Moncrief,  Phys. Rev. D {\bf{48}} 4676 (1993); B. K. Berger, Living Rev. Rel. 
{\bf{5}} 1 (2002).


\bibitem{DavidG}
D. Garfinkle, Phys. Rev. Lett. {\bf{93}} 161101 (2004);
D. Garfinkle,   Class. Quant. Grav. {\bf{24}} S295 (2007).



\bibitem{WE}
J. Wainwright and G. F. R. Ellis, ``Dynamical systems in cosmology'' (Cambridge
University Press, Cambridge, 1997).


\bibitem{Uggla03} C. Uggla, H. van Elst, J. Wainwright and G. F. R. Ellis, 
%%The past attractor in inhomogeneous cosmology. 
Phys. Rev. D {\bf{68}} 103502 (2003).



\bibitem{Heinzle}
J. M. Heinzle and C. Uggla,  
%%Mixmaster: Fact and Belief, 
Class. Quant. Grav. {\bf{26}}
075016 (2009) [arXiv:0901.0776].

\bibitem{Rendall}
A. D. Rendall,
%%Global dynamics of  the  mixmaster model. 
Class. Quant. Grav. {\bf{14}}
2341 (1997).

\bibitem{Ringstrom}
H. Ringstrom,
%% Curvature blow up in Bianchi VIII and IX vacuum spacetimes. 
Class.
Quant. Grav. {\bf{17}} 713 (2000): H. Ringstrom,
%%The Bianchi IX attractor, 
Annales Henri Poincare {\bf{2}} 405 (2001).

\bibitem{Brehm}
B. Brehm, ``Bianchi VIII and IX vacuum cosmologies: Almost every so-
lution forms particle horizons and converges to  the  Mixmaster attractor''
(Doctoral Thesis, Freie Universitat Berlin, 2016)
[arXiv:1606.08058, 2016].



\bibitem{Heinzle09}
J. M. Heinzle, C. Uggla and N. Rohr, 
%%The cosmological billiard attractor. 
Adv. Theor.
Math. Phys. {\bf{13}} 293 (2009).

\bibitem{BGZK} V. A.
Belinskii, L. P. Grishchuk, Ya. B. Zeldovich and I. M.  Khalatnikov,
%%Inflationary stages in cosmological models with a scalar field,
Sov. Phys. JETP {\bf{62}} 195 (1986).


\bibitem{Hewitt2003} C. G. Hewitt, J. T. Horwood and J. Wainwright,
%%Asymptotic dynamics of  the  exceptional Bianchi cosmologies,
 Class. Quant. Grav. {\bf{20}} 1743 (2003).


\bibitem{LeBlanc} V. G.
LeBlanc,
%% Asymptotic states of magnetic Bianchi I cosmologies,
Class. Quant. Grav. {\bf{14}} 2281 (1997); V. G.
LeBlanc,  D. Kerr, and J. Wainwright, 
%%Asymptotic states of magnetic Bianchi VI0 cosmologies, 
Class. Quant. Grav. {\bf{12}} 513
(1995).

\bibitem{Hewitt2001}
 C. G. Hewitt, R. Bridson and  J. Wainwright, 
%%The asymptotic regimes of tilted Bianchi II cosmologies, 
Gen. Rel. Grav. {\bf{33}}  65 (2001). 


\bibitem{Hervik07}
S. Hervik, R. J. van den Hoogen, W. C. Lim and A. A. Coley, 
%%Late-time behaviour of  the  tilted Bianchi type VIh models: 
Class. Quant. Grav. 
{\bf{24}} 3859 (2007) [arXiv:gr-qc/0703038].



\bibitem{Uggla13} 
C. Uggla,
%% Spacetime Singularities: Recent Developments
Int. J. Mod. Phys. D {\bf{22}} 1330002 (2013) [arXiv:1306.6527] \& 
 %%Recent developments concerning generic spacelike singularities.
Gen. Rel. Grav. {\bf{45}} 1669 (2013) [arXiv:1304.6905].




\bibitem{Maier} R. Maier, I. Damiao Soares and E. V. Tonini,
%%General Bianchi IX dynamics in bouncing braneworld cosmology: homoclinic chaos and  the  BKL conjecture, 
Class. Quant. Grav. {\bf{32}} 235001 (2015) 
[arXiv:1505.06189].

\bibitem{Temam}
R. Temam,  ``Infinite-Dimensional Dynamical Systems in Mechanics
and Physics'' (Springer-Verlag, 1988); 
R.  Temam,  ``Dynamical systems in infinite dimensions'', in ``The Connection
between Infinite-Dimensional and Finite-Dimensional Dynamical
Systems'', eds. B. Nicolaenko, C. Foias and R. Temam (American
Mathematical Society, 1988).


\bibitem{Damour} T. Damour, M. Henneaux and H. Nicolai,
%% Cosmological billiards. 
Class. Quant.
Grav. {\bf{20}} R145 (2003) [hep-th/0212256]; 
T. Damour, M. Henneaux and H. Nicolai, 
%%“E(10) and a ’small tension expansion’ of M theory,”
Phys. Rev. Lett. {\bf{89}} 221601 (2002) [hep-th/0207267];
see also  T. Damour and H. Nicolai, 
%%Higher order M theory corrections and  the  Kac-Moody algebra E10, 
Class. Quant. Grav. {\bf{22}} 2849 (2005).


\bibitem{Kleinschmidt}
T. Damour and C. Hillmann, 
%%Fermionic Kac-Moody Billiards and Supergravity, 
JHEP {\bf{0908}} 100 (2009) [arXiv:0906.3116 [hep-th]];
A. Kleinschmidt, M. Koehn and H. Nicolai, 
%%Supersymmetric quantumcosmological billiards, 
Phys. Rev. D {\bf{80}} 061701 (2009) [arXiv:0907.3048 [gr-qc]].

\bibitem{DamourSpindel} T. Damour and P. Spindel, 
%%Quantum supersymmetric cosmology and its hidden Kac-Moody structure, 
Class. Quant. Grav. {\bf{30}}  162001 (2013) [arXiv:1304.6381[gr-qc]] \&
%%“Quantum Supersymmetric Bianchi IX Cosmology,” 
Phys. Rev. D{\bf{ 9}} 103509 (2014) [arXiv:1406.1309 [gr-qc]] \& Phys. Rev. D {\bf{95}} 126011 (2017) 
[arXiv:1704.08116].
%%Quantum Supersymmetric Cosmological Billiards and their Hidden Kac-Moody Structure DamourSpindel 


\bibitem{RodnianskiSpeck}
I. Rodnianski and J. Speck,
``Stable Big Bang Formation in Near-FLRW Solutions to  the  Einstein-Scalar Field and Einstein-Stiff Fluid 
Systems''   [arxiv.org/1407.6298].
%% -- no ref.


\bibitem{RS3}
I. Rodnianski and J.  Speck, ``On  the  nature of Hawking's incompleteness for  the  Einstein-vacuum 
equations'' [arxiv:1804.06825].
%% -- no ref.
 

\bibitem{RodnianskiSpeck2}
I. Rodnianski and J.  Speck,
``The Stability of  the  Irrotational Euler-Einstein System with a Positive Cosmological Constant''
[arxiv/0911.5501].
%% -- no ref.
%%





\bibitem{Ringstrom2004}
H. Ringstrom,  Math. Proc. Camb. Phil. Soc. {\bf{136}} 485 (2004);
H. Ringstrom,  Class. Quant. Grav. {\bf{21}} S305 (2004).




\bibitem{art:Lim2015}
W.~C. Lim, 
Class. Quant. Grav. {\bf 32} 162001 (2015) [arXiv:1507.02754]; A. A. Coley, D. Gregoris and W. C. Lim, 
Class. Quant. Grav. (2016) [arXiv:1606.07177].


\bibitem{art:Limetal2009}
W.~C. Lim, L.~Andersson, D.~Garfinkle and F.~Pretorius,
Phys. Rev. D {\bf 79} 103526 (2009) [arXiv:0904.1546].


  
\bibitem{art:Bergeretal2001}
B.~K. Berger, J.~Isenberg and M.~Weaver, 
Phys. Rev. D {\bf 64} 084006 (2001).

\bibitem{art:ColeyLim2012}
A.~A. Coley and W.~C. Lim, 
Phys. Rev. Lett. {\bf 108} 191101 (2012) [arXiv:1205.2142]; W.~C. Lim and A.~A. Coley,
Class. Quant. Grav. {\bf {31}} 015020 (2014) [arXiv:1311.1857].

\bibitem{HeinzleUggla2012}
J. M. Heinzle and  C. Uggla,
Gen. Rel. Grav.  {\bf{45}}  939 (2013) [arXiv:1212.5500/gr-qc];
J. M. Heinzle, C. Uggla and W. C. Lim, 
Phys. Rev. D {\bf{86 }}    104049 (2012) [arXiv:1206.0932/gr-qc].   


\bibitem{Computational}
A.~Coley, L. Lehner, F. Pretorius and D. ~Wiltshire,
``Computational Issues in Mathematical Cosmology'' (2017);
http://cms.iopscience.iop.org/alfresco/d/d/workspace/SpacesStore/
83f10d6e-0b33-11e7-9a47-19ee90157113/Overview-CC.pdf


\bibitem{Bentivegna}
E. Bentivegna and
M. Bruni,  Phys.  Rev.  Lett.  {\bf{116}} 251302 (2016) [arXiv:1511.05124];
E. Bentivegna, Phys.
Rev. D {\bf{95}} 044046 (2017) [arXiv:1610.05198];
J. T.  Giblin, J. B.
Mertens and G. D. Starkman,  Phys.  Rev.  Lett.  {\bf{ 116}} 251301 (2016),
{\em{ibid.}} Phys.  Rev.  D {\bf{93}} 124059 (2016) [arXiv:1511.01105],
{\em{ibid.}} ``A cosmologically motivated reference formulation of numerical relativity''
[arXiv:1704.04307];
J. Adamek, D. Daverio, R. Durrer and M. Kunz,  Nature
Physics {\bf{12}} 346 (2016) [arXiv:1509.01699];
J. Adamek, C. Clarkson, D. Daverio, R. Durrer and M. Kunz,
``Safely smoothing spacetime: backreaction in relativistic cosmological simulations''
[arXiv:1706.09309];
H. Macpherson, D. J. Price and P. D. Lasky, 
``Einstein's Universe: cosmological structure formation in numerical relativity''
[arXiv:1807.01711].




\bibitem{refluis}
J. R. Westernacher-Schneider, L. Lehner and Y. Oz,  JHEP {\bf{12}} 067 (2015)
[arXiv:1510.00736].
%%Scaling Relations in Two-Dimensional Relativistic Hydrodynamic Turbulence

\bibitem{Barack} 
L. Barack {\em{et al.}},
``Black holes, gravitational waves and fundamental physics: a roadmap''
[arXiv:1806.05195].

\end{thebibliography}
\end{document}